\documentclass[conference]{IEEEtran}
\usepackage{fancyhdr}

\usepackage[style=ieee]{biblatex}
\usepackage{amsmath,amssymb,amsfonts}
\usepackage{graphicx}
\usepackage{textcomp}

\usepackage[utf8]{inputenc}
\usepackage[english]{babel}
\usepackage{booktabs}
\usepackage[hidelinks]{hyperref}
\usepackage{nowidow}
\usepackage{bm}
\usepackage[dvipsnames]{xcolor}
\usepackage{algorithm}
\usepackage{algpseudocode}
\usepackage{mathrsfs}

\usepackage{makecell} 

\newcommand*{\mat}[1]{\ensuremath{\bm{#1}}}
\newcommand*{\sign}{\ensuremath{\text{sign}}}
\newcommand*{\Tr}{\ensuremath{\text{Tr}}}

\begin{document}

\title{{ Towards} Electronic Structure-Based \textit{Ab-Initio} Molecular Dynamics Simulations with Hundreds \\ of Millions of Atoms}

\author{Robert Schade\IEEEauthorrefmark{1}, Tobias Kenter\IEEEauthorrefmark{1}\IEEEauthorrefmark{2}, Hossam Elgabarty\IEEEauthorrefmark{3}, \IEEEauthorblockN{Michael Lass\IEEEauthorrefmark{1}\IEEEauthorrefmark{2}, Ole Sch\"utt\IEEEauthorrefmark{5}, \\ Alfio Lazzaro\IEEEauthorrefmark{6}, Hans Pabst\IEEEauthorrefmark{7}, Stephan Mohr\IEEEauthorrefmark{8}\IEEEauthorrefmark{9}, Jürg Hutter\IEEEauthorrefmark{10}, Thomas D. Kühne\IEEEauthorrefmark{1}\IEEEauthorrefmark{3}\IEEEauthorrefmark{4}, Christian Plessl\IEEEauthorrefmark{1}\IEEEauthorrefmark{2}}
%
\IEEEauthorblockA{Paderborn University, Warburger Str. 100, 33098 Paderborn, Germany}
\IEEEauthorblockA{\IEEEauthorrefmark{1}Paderborn Center for Parallel Computing \hspace{1ex} \IEEEauthorrefmark{2}Department of Computer Science}
\IEEEauthorblockA{\IEEEauthorrefmark{3}Department of Chemistry \hspace{1ex} \IEEEauthorrefmark{4}Corresponding author, Email: thomas.kuehne@uni-paderborn.de}
%

\IEEEauthorrefmark{5}Department of Materials, ETH Zürich, CH-8092 Zürich, Switzerland \\
\IEEEauthorrefmark{6}HPE Switzerland GmbH, Basel, Switzerland \\
\IEEEauthorrefmark{7}Intel Extreme Computing, Software and Systems, Zürich, Switzerland \\
\IEEEauthorrefmark{8}Nextmol (Bytelab Solutions SL), Barcelona, Spain \hspace{1ex}
\IEEEauthorrefmark{9}Barcelona Supercomputing Center (BSC) \\
\IEEEauthorrefmark{10}Department of Chemistry, University of Zurich
}
\maketitle
\thispagestyle{fancy}
\lhead{}
\rhead{}
\chead{}
\rfoot{}
\cfoot{}
\renewcommand{\headrulewidth}{0pt}
\renewcommand{\footrulewidth}{0pt}

\begin{abstract}
    
   We push the boundaries of electronic structure-based \textit{ab-initio} molecular dynamics (AIMD) beyond 100 million atoms. This scale is otherwise barely reachable with classical force-field methods or  novel  neural  network  and  machine  learning  potentials. We achieve this breakthrough by combining innovations in linear-scaling AIMD, efficient and approximate sparse linear algebra, low and mixed-precision floating-point computation on GPUs, and a compensation scheme for the errors introduced by numerical approximations.
   
    The core of our work is the non-orthogonalized local submatrix method (NOLSM), which scales very favorably to massively parallel computing systems and translates large sparse matrix operations into highly parallel, dense matrix operations that are ideally suited to hardware accelerators. We demonstrate that the NOLSM method, which is at the center point of each AIMD step, is able to achieve a sustained performance of 324 PFLOP/s in mixed FP16/FP32 precision corresponding to an efficiency of 67.7\% when running on 1536 NVIDIA A100 GPUs.
    \end{abstract}


\section{Overview of the Problem}

\subsection{Atomistic Computer Simulations}
The exponential increase in the performance of high-performance computers over the past decades, together with advances in computer science and applied mathematics, has led to the birth of a new way of doing science at the intersection of theory and experiment. This field is generally referred to as computational science and allows for experiments \textit{in silico} that otherwise would be too difficult, expensive, or simply impossible to perform. As a result, computer simulations have been very successful in predicting and rationalizing a large variety of novel physical phenomena. 

For systems made of atoms, the two most common computational techniques to conduct such simulations are the Monte Carlo and the molecular dynamics (MD) algorithms \cite{MC1953,PhysRev.136.A405}. The latter is simply the numerical solution of Hamilton’s equation of motion, which allows both equilibrium thermodynamic and dynamic properties of a system at finite temperature to be computed. Since it also provides a window into the real-time evolution of the atoms, another role of MD is that of a computational microscope.

One of the most challenging and very important aspects of MD simulations is calculating the interatomic forces. In classical simulations they are computed by conventional force fields, or novel neural network and machine learning potentials, which have been parameterized to reproduce experimental or accurate \textit{ab-initio} data of small model systems \cite{UFF, Karplus}. Even though great strides in improving such empirical potentials have been made and often render them surprisingly accurate \cite{Ceriotti, Tkatchenko}, the transferability to systems or regions of the phase diagram different from the ones to which they have been trained in the first place may be restricted. 
Ultimately, when assuming a classical model, as ingenious it may be, the access to the quantum mechanical electronic structure is irrevocably lost. However, some of the most relevant and interesting phenomena of modern chemistry and physics are inherently non-classical.

\subsection{Electronic Structure-based \textit{Ab-initio} Molecular Dynamics}
Therefore, an electronic structure-based \textit{ab-initio} MD (AIMD) approach \cite{PhysRevLett.55.2471, RevModPhys.64.1045}, where the forces are computed on-the-fly from accurate quantum mechanical calculations, is very attractive since many of these limitations can, in principle, be removed. Nevertheless, the accuracy and increased predictive power of AIMD simulations come at a significant computational cost, which has to be carefully balanced against system size and sampling requirements, thus limiting the attainable length and time scales despite substantial progress \cite{PhysRevLett.98.066401}. Hence, effective single-particle theories such as Hartree-Fock, density functional theory (DFT), and semi-empirical tight-binding (TB) approaches are to date the most commonly used electronic structure methods within AIMD \cite{WIREs}. However, for very large systems, like those occurring in biology, nanotechnology, materials science, or mechanical engineering, which contain many millions of atoms, self-consistently solving the corresponding Schr\"odinger-like equations is computationally not feasible even on today's largest supercomputers. This practical limitation stems from the fact that these equations are very high-dimensional eigenvalue problems with up to trillions of unknowns. The computation of all eigenvalues and corresponding eigenvectors requires the diagonalization of the quantum mechanical Hamilton operator that uniquely defines the specific system and typically scales cubically with its size.

\subsection{Linear-scaling Electronic Structure Theory}
Therefore, novel computational methods that scale linearly with the size of the system to directly calculate the all-important density matrix instead of all eigenvectors would be very desirable, thus making a new class of systems accessible to AIMD that were previously thought not feasible. 
Several so-called linear-scaling methods have been proposed to circumvent the cubic scaling diagonalization that is the main bottleneck of DFT and TB \cite{RevModPhys.71.1085,PhysRevLett.66.1438,PhysRevLett.69.3547,Richters}. Underlying all of these methods is the concept of ``nearsightedness'' \cite{ProdanKohn}, an intrinsic system-dependent property, which states that at fixed chemical potential the electronic density depends just locally on the external potential so that all matrices required to compute the Fermi operator will become sparse \cite{kuhne2020disordered}. When using sparse matrix algebra techniques the property can be exploited to devise computational methods whose memory requirements and computational effort increase only linearly with problem size. However, the crossover point after which linear-scaling electronic structure methods become advantageous has remained rather large, in particular if high accuracy is needed.

\subsection{Approximate Computing-Based Submatrix Method}
Beyond algorithmic improvements, it is also possible to relax the requirement for the accuracy of computations and profit from the substantially improved performance of modern computer architectures for low-precision arithmetic.
We demonstrate that by leveraging the approximate computing (AC) paradigm \cite{AC1,AC2}, the usage of mixed- and low-precision numerics can be rigorously compensated by an appropriately modified Langevin-type equation. The noise within the nuclear forces can be assumed as white, thus facilitating the exact computation of ensemble-averaged expectation values \cite{Richters, karhan2014role, Computation}. One possible route is the linear-scaling sign-method \cite{sign-method}, {  whose accuracy had been previously systematically investigated in detail and it has been demonstrated that good accuracy can be obtained \cite{SM}. The chief advantage of the employed sign-method, however, is that it only relies on large sparse matrix-matrix multiplies. Yet,} due to the distributed nature of the required matrix multiplications, large-scale applications are usually limited by a communication bottleneck. To avoid this bottleneck, we have extended the recently developed submatrix method \cite{Lass_Mohr_Wiebeler_Kühne_Plessl_2018, Lass_Schade_Kühne_Plessl_2020}, which transforms calculations on large distributed sparse matrices into computations on small local dense matrices and combined it with the second-generation Car-Parrinello method of K\"uhne et al. \cite{PhysRevLett.98.066401,kuhne2018disordered} to bypass the previously mentioned self-consistent solution. This transformation opens the door to employ hardware-accelerated low-precision linear algebra without compromising the accuracy of the eventual results.

\section{Current State of the Art}

Previous attempts 
to push the boundaries of electronic structure-based structure relaxation and AIMD simulations are summarized in Table~\ref{PrevRecords}.
\begin{table*}[h!]
\centering
\caption{\label{PrevRecords} Performance of previously conducted electronic structure-based structure relaxation or AIMD simulations. Therein, the employed electronic structure method is abbreviated by DFT, NSC-DFT, LS-DFT and SS-DFT, which stands for density functional theory and its non-self-consistent, linear-scaling and subsystem variants, respectively. The corresponding basis set to represent the single-particle orbitals are denoted by PW for conventional plane waves, RMG-PW for real-space multigrid plane waves, GPW for Gaussian and plane waves, GTO for Gaussian-type orbitals, FD for finite difference, RS-FD for real-space finite difference, FEM for finite element method, NGWF for non-orthogonal generalized Wannier functions and PAO for polarized atomic orbitals. If the calculation was conducted involving trivial k-point parallelism, the total number of atoms is given as the product of number of independent instances time the number of atoms in anyone of them. The sustained efficiency is either given with respect to the corresponding peak performance, or estimated in terms of parallel efficiency and identified by the ``$\approx$'' sign.}
\begin{tabular}{|l|c|c|c|c|c|c|c|c|c|c|}
     \hline
     \textbf{Code} & \textbf{Year} & \textbf{Method} & \textbf{Basis} & \textbf{System} & \textbf{\# Atoms} & \textbf{\# Cores} & \textbf{Machine} & \textbf{\makecell{Peak\\Performance}} & \textbf{Efficiency} \\ \hline \hline
     CPMD \cite{hutter2005dual} & 2005 & DFT & PW & bulk SiC & 1k & 1.2k CPU & IBM p690 & 1.087 TFLOP/s & $\approx$ 20\% \\ \hline
     Qbox \cite{gygi2006large} & 2006 & DFT & PW & bulk Mo & 8*1k & 128k CPU & IBM BlueGene/L & 207.3 TFLOP/s & 56.5\% \\ \hline
     LS3DF \cite{zhao2009linearly} & 2009 & DFT & PW & bulk ZnTeO & 36k & 147k CPU & Cray Jaguar & 442 TFLOP/s & $\approx$ 33\% \\ \hline
     CONQUEST \cite{bowler2010calculations} & 2010 & NSC-DFT & PAO & bulk Si & 2.1M & 4k CPU & Cray XT4 & & $\approx$ 60\% \\ \hline
     CP2K \cite{vandevondele2012linear} & 2012 & LS-DFT & GPW & bulk H$_2$ & 1M & 47k CPU & Cray XT5 & & \\ \hline
     ONETEP \cite{wilkinson2014hybrid} & 2014 & LS-DFT & NGWF & {\makecell{amyloid fibril\\ trimer}} & 42k & 115k CPU & IBM BlueGene/Q & & \\ \hline
     CONQUEST \cite{arita2014large} & 2014 & LS-DFT & PAO & bulk Si & 786k & 200k CPU & K-Computer & & \\ \hline
     RSDFT \cite{hasegawa2014performance} & 2014 & DFT & RS-FD & Si nanowire & 107k & 664k CPU & K-Computer & 5.48 PFLOP/s & 51.67\% \\ \hline
     CP2K \cite{andermatt2016combining} & 2016 & SS-DFT & GPW & {\makecell{satellite tobacco \\ mosaic virus}}  & 1M & 20k CPU & Cray XC30 & & \\ \hline
     LDC-DFT \cite{nomura2014metascalable} & 2014 & SS-DFT & RMG-PW & bulk SiC & 6.3M & 786k CPU & IBM BlueGene/Q & 5.08 PFLOP/s & 50.5\% \\ \hline
     OpenAtom \cite{jain2016openatom} & 2016 & DFT & PW & periodic MOF & 32*424 & 262k CPU & IBM BlueGene/Q & & $\approx$ 52\% \\ \hline
     MGmol \cite{fattebert2016modeling} & 2016 & LS-DFT & FD & bulk H$_2$O & 1.2M & 1.6m CPU & IBM BlueGene/Q & & $\approx$ 39\% \\ \hline
     DFT-FE \cite{das2019fast} & 2019 & DFT & FEM & Mg cluster & 10.5k & \makecell{159k CPU \\ +22.8k GPUs} & IBM Summit & 46 PFLOP/s & 27.8\% \\ \hline
     This work & 2021 & LS-DFT & GTO & bulk water & 102M & \makecell{18.4k CPU\\ +1.5k GPUs} & JUWELS Booster & 206 PFLOP/s & 43\% \\ \hline
     This work & 2021 & LS-DFT & GTO & {\makecell{HIV-1 capsid \\ in solution}} & 62.5M & \makecell{18.4k CPU \\ +1.5k GPUs} & JUWELS Booster & 324 PFLOP/s & 67.7\% \\ \hline
\end{tabular}
\end{table*}
They include DFT calculations using delocalized plane wave (PW) basis sets, as implemented in the CPMD \cite{hutter2005car}, Qbox \cite{gygi2008architecture}, LS3DF \cite{zhao2009linearly} and OpenAtom \cite{jain2016openatom} codes, as well as localized orbital DFT simulations based on real-space finite difference (RS-DFT) and finite element methods (FEM) using the RSDFT \cite{hasegawa2014performance} and DFT-FE \cite{motamarri2020dft} codes, respectively. The largest simulations, however, are conducted employing low-scaling electronic structure methods such as linear-scaling DFT (LS-DFT). With the exception of the LDC-DFT code \cite{nomura2014metascalable}, which relies on an extended real-space multigrid PW (RMG-PW) basis within a less correlated subsystem DFT (SS-DFT) approach, localized basis functions with finite spatial extent are used. Examples of the latter are non-orthogonal generalized Wannier functions (NGWF), finite difference (FD), polarized atomic orbitals (PAO) and Gaussian and plane waves (GPW) basis sets that are implemented in the ONETEP \cite{prentice2020onetep}, MGmol \cite{fattebert2016modeling}, CONQUEST \cite{nakata2020large} and CP2K \cite{CP2K} codes, respectively. 

To put our achievement in context to previous work,  we find it important to point out that the hitherto largest electronic structure calculation conducted so far with 6.3 million atoms has been achieved using the SS-DFT approach, which subdivides the total system in a divide-and-conquer fashion into overlapping fragments that can be computed independently from each other. Even though this offers an intriguing additional level of parallelism, which is reflected by a peak performance of more than 5~PFLOP/s, it also entails a further approximation. Along similar lines, simulations involving multiple independent k-points can also be trivially parallelized over each of these points. Interestingly, the simulation with the so-far largest peak performance of 46~PFLOP/s and an efficiency of 27.8\% has been conducted for just 10.5 thousand atoms, even though with electronically rather complicated alkaline earth metals atoms. 

In the present work, we have conducted individual AIMD-based dynamical simulated-annealing steps to mimic the relaxation of the structure of a whole human immunodeficiency virus-1 (HIV-1) capsid in aqueous solution containing more than 62.5 million atoms, as well as for water with about 102 million atoms. For that purpose, we have extended the Geometry, Frequency, Noncovalent, eXtended Tight-Binding (GFN-xTB) scheme towards periodic systems and implemented it within the CP2K code \cite{doi:10.1021/acs.jctc.7b00118,CP2K}.

\section{Innovations Realized}

\subsection{Summary of Contributions}
The central innovation of this work is the approximate mapping of a matrix function of a very large sparse matrix to a series of matrix functions of much smaller but dense matrices. 
Since in this way inter-node communication is avoided, a very favorable parallel scaling is obtained. 
The evaluations of the matrix functions for the small dense matrices, with dimension of $\sim500$ to $\sim 10000$ for the applications in this work, are computed with iterative schemes and mixed-precision arithmetic on tensor cores of GPUs. The resulting noise from these approximations is rigorously compensated by making use of the fluctuation-dissipation theorem so that the desired thermodynamic expectation values can nevertheless be obtained accurately.

\subsection{Algorithmic Innovations}
\subsubsection{Approximate Computing}
The ideas of AC can be applied to the field of electronic structure-based AIMD simulations by recognizing that algorithmic or numerical approximations cause noise in the computed total energy $E^N$ of the system and in consequence noise $\mat  \Xi_i$ in the forces $\mat F_i$ on the atoms. Thus, the computed noisy forces $\mat F^N_i$ can be written as 
\begin{equation}
    \mat F^N_i=-\frac{\partial E^N}{\partial \mat  R_i}=\mat  F_i+\mat \Xi_i, \label{eq:Eforces}
\end{equation}
where $\mat F_i$ denote the exact forces. All quantities depend on the position of the atoms $\mat R_1,...,\mat  R_n$.
In previous works we have demonstrated that in the present context $\mat \Xi_i$ can be assumed to be nearly unbiased \cite{Richters, karhan2014role, Computation}, thus fulfilling the so-called fluctuation-dissipation theorem
\begin{equation}
    \langle \mat \Xi_i(0) \cdot\mat \Xi_i(t)\rangle_T \approxeq 2 \gamma_N M_i k_B T \delta(t),
    \label{FDT}
\end{equation}
where $\langle ... \rangle_T$ denotes the Boltzmann-weighted ensemble average at the temperature $T$, $k_B$ the Boltzmann constant, $M_i$ the atomic masses, and $\gamma_N$ a friction coefficient, whose exact value needs to be determined. 
However, if we would know $\gamma_N$ such that Eq.~\ref{FDT} is satisfied, a modified Langevin-type equation 
\begin{equation}
    M_i \ddot {\mat R}_i=\mat F_i+ \mat  \Xi_i-\gamma_N M_i \dot {\mat R}_i
    \label{eq:ModLangevinEq}
\end{equation}
is recovered, which guarantees for an accurate canonical sampling of the Boltzmann distribution and to compute precise thermodynamic expectation values \cite{PhysRevLett.98.066401}.
Fortunately, the exact value of $\gamma_N$ does not need to be known \textit{a priori}, but can be bootstrapped so as to generate the correct average temperature \cite{kuhne2018disordered}, as measured by the equipartition theorem 
{ 
\begin{equation}
    \left\langle \frac{1}{2} M_i \dot {\mat R}_i^2 \right\rangle = \frac{3}{2} k_B T. \label{Equipartition}
\end{equation}
More precisely, in order to determine the hitherto unknown value of $\gamma_N$, we perform a short preliminary simulation on an identical but smaller system in which we vary $\gamma_N$ on-the-fly using a Berendsen-like algorithm until Eq.~\ref{Equipartition} is eventually satisfied \cite{WaterAIMD}. Previous studies have demonstrated the efficacy of this approach for a wide variety of systems ranging from insulators to semiconductors and even to metals in condensed phases \cite{WIREs,Computation}.}
 
\subsubsection{Linear-Scaling Eigenvalue Solver via the Non-Orthogonalized Local Submatrix Method}\label{sec:localsubmatrix}
\paragraph{Linear-Scaling Electronic Structure Calculations}
In electronic structure-based AIMD simulations, forces $\mat F_i$ on the atoms are derived in every time step on-the-fly from the solution of the quantum mechanical problem of electrons in the electrostatic field generated by the nuclei. The total energy of a system can be written as
\begin{equation}\label{eq:etot}
    E=E_{elec}+ E_{dc} + E_{ion}=\sum_i^{occ.} \langle \psi_i|\hat H_0|\psi_i\rangle+E_{dc}+E_{ion},
\end{equation}
where the summation runs over all occupied electronic states $|\Psi_i\rangle$ in the ground state, the Hamiltonian operator $\hat H_0$, additional double counting terms $E_{dc}$ and the nuclear Coulomb repulsion energy $E_{ion}$. The form of the Hamiltonian matrix $\mat H_0$ and the double counting terms depend on the level of the theory. 
In any case, however, $\mat H_0$ is dependent on the one-particle density matrix $\mat D$, or on the electron density, which necessitates a self-consistency cycle (SCF). 
A linear-scaling algorithm to solve the quantum mechanical problem, $\hat H_0|\psi_i\rangle=\epsilon_i |\psi \rangle$, is required to find the ground-state energy of the system that determines the forces via Eq.~\ref{eq:Eforces}. Linear-scaling density-matrix-based electronic structure algorithms directly purify the Hamiltonian into the density matrix $\mat D$ \cite{mcweeny1960some}, i.e.,
\begin{equation}
    \mat D=\frac{1}{2}\left(\mat I-\sign(\mat S^{-1}\mat H_0-\mu \mat I)\right)\mat S^{-1}, \label{eq:purification}
\end{equation} 
where $\mat S$ denotes the overlap matrix and $\mu$ the chemical potential. The electronic energy and the forces can now be obtained via
\begin{equation}\label{eq:e_elec}
    E_{elec}=\Tr(\mat D\mat H_0).
\end{equation}
To evaluate the contribution to the forces from the localized atom-centered basis functions (Pulay forces) \cite{pulay1969ab}, the energy-weighted density matrix
\begin{equation}
    \mat W=\mat D\mat H_0\mat D
\end{equation} 
is required.
The matrix-sign function in Eq.~\ref{eq:purification} can be evaluated iteratively, for example with the Newton-Schulz iteration \cite{doi:10.1002/zamm.19330130111}
\begin{align}\label{eq:NS1}
    \mat X_0&=\mat A, \ \ \mat X_{k+1}=\frac{1}{2}\mat X_k (3\mat I-\mat X_k^2)\\ \label{eq:NS2}
    \sign(\mat A)&=\lim_{k\rightarrow \infty} \mat X_k,
\end{align}
 which converges quadratically if the matrix $\mat A$ has no purely imaginary eigenvalues and $\|\mat A^2-I\|<1$~\cite{Kenney1991-sa}. The matrices $\mat A$ in this work have a real eigenspectrum and the probability that an eigenvalue is numerically zero is negligible. The norm condition is ensured by a rescaling of the matrix prior to the sign iteration. Other possibilities than the Newton-Schulz iteration are \color{black} higher-order Padé-approximants \cite{Higham1997}, or arbitrary-order iteration schemes \cite{Richters_Lass_Walther_Plessl_Kühne_2019}  . In conventional linear-sclaing schemes the \color{black} underlying multiplications of large sparse matrices are performed with global matrix operations \cite{sign-method,CP2K}.
In contrast, we view the purification as a matrix function and approximate it with our submatrix method so that no global matrix multiplications are required that would otherwise lead to a communication-bound algorithm.

\begin{figure}
    \centering
    \includegraphics[width=0.475\textwidth]{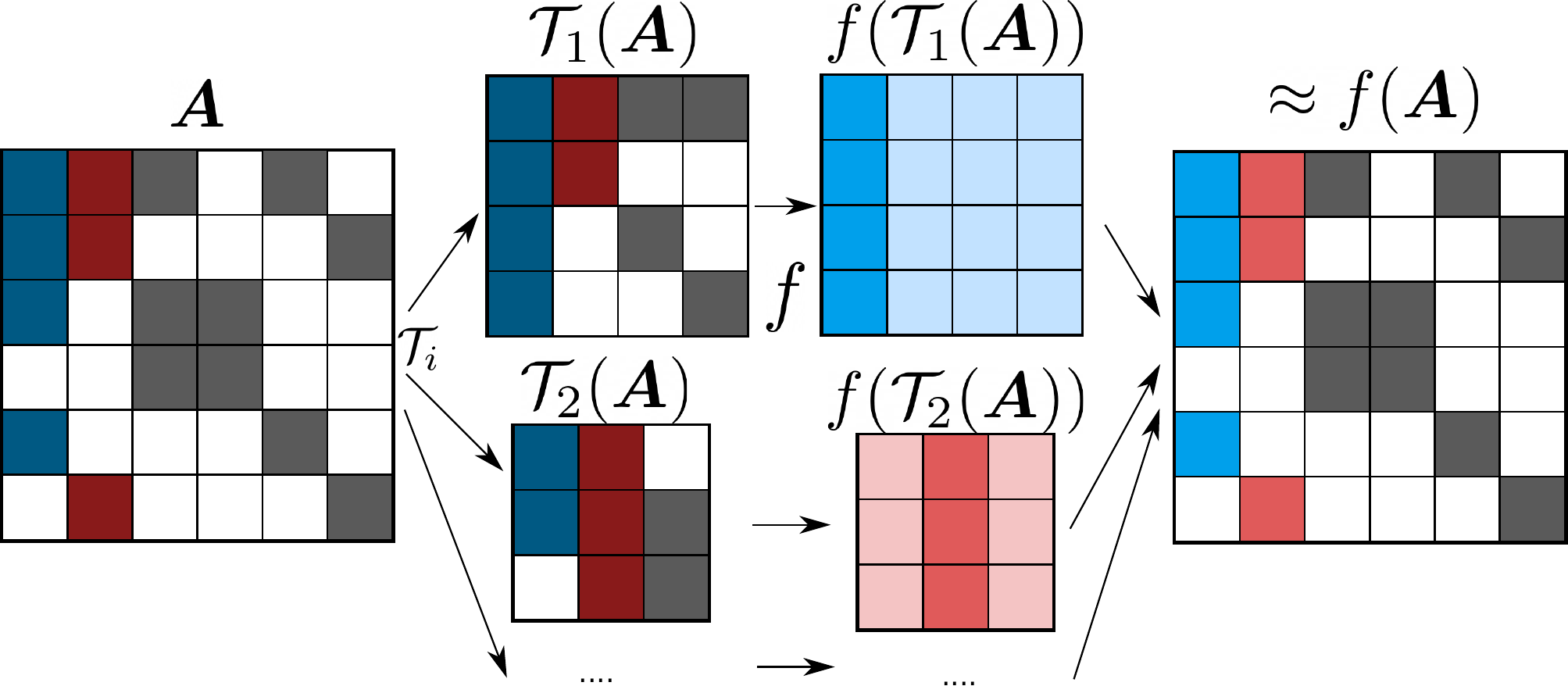}
    \caption{Schematic representation of the steps of the submatrix method for the approximate calculation of a matrix function $f(\mat A)$ of a large sparse matrix $\mat A$. The first step is the construction of a submatrix $\mathcal T_i(\mat A)$ for every column of the matrix $\mat A$. Then the matrix function is applied to the dense submatrices, i.e., $f(\mathcal T_i(A))$ and finally the relevant result columns are inserted into the sparse result matrix.}
    \label{fig:submatrixgeneral}
\end{figure}

\paragraph{The Submatrix Method}\label{sec:submatrix_general}
The submatrix method \cite{Lass_Mohr_Wiebeler_Kühne_Plessl_2018, Lass_Schade_Kühne_Plessl_2020}, recently developed by some of the authors, approximates a matrix function of a large sparse matrix by evaluating it on a series of much smaller and denser matrices. The underlying idea of the submatrix method is described by Fig.~\ref{fig:submatrixgeneral}.   It entails three major steps: (i) submatrix construction, (ii) application of the matrix function and (iii) write-back of the result. In the first step a submatrix is constructed for each column of the original matrix $\mat A$ by removing the rows and their corresponding columns, where the chosen column has zero values. Hence, \color{black} $\mathcal T_j$ constructs the submatrix for column $j$. Thus, $\mathcal T_j(\mat A)$ represents the dense submatrix that is constructed for column $j$ of the large sparse matrix $\mat A$.
  
In the second step the matrix function $f$ is applied to every submatrix independently, i.e., $f(\mathcal T_j(\mat A))$ for every $j$. Because the submatrices are by construction much denser and much smaller than the original matrix, efficient dense matrix algorithms can be employed for all operations. 
In the last step, the write-back step, the relevant column of the matrix $f(\mathcal T_j(\mat A))$ is written back into the $j$-th column of the result matrix and represents an approximation of the $j$-th column of $f(\mat A)$. 

As a result, the submatrix approximation of $f(\mat A)$ has the same sparsity pattern as the matrix $A$. Please note, that the submatrix merging procedure proposed in Sec.~\ref{sec:submatrix_combination} can make the submatrix approximation of $f(\mat A)$ more dense than $\mat A$. Moreover, when applied to well behaved matrices, i.e., not arrowhead-type matrices the
\color{black}
dimensionality of the submatrices are independent of the system size for a sufficiently large system (i.e. within the linear-scaling regime).   This is for example the case if the submatrix method is applied to the overlap, or Hamiltonian matrix and when localized basis functions are used to describe the electronic wave functions.

The application of the submatrix method to an atomistic system described with local atom-centered basis functions can be understood by identifying the columns (or sets of columns) of the input matrix $\mat A$ with atoms. The construction of a submatrix for a column $j$ corresponding to an atom $J$ can then be seen as the the construction of a subsystem containing all atoms in the vicinity of the atom $J$ that have non-zero matrix elements with atom $J$. For this subsystem, the density matrix is computed as a matrix function and the matrix elements between the atom $J$ and other atoms are used as an approximation to the elements of the density matrix of the full system.
Thus, the submatrix method makes use of the intrinsic nearsightedness of the electronic matter and adaptively defines an environment around each atom as a subsystem \cite{ProdanKohn}.

Hence, the submatrix method is particularly suitable to estimate the total energy and quantities derived from it. The electronic energy $E_\mathrm{elec}$ given in Eq.~\ref{eq:e_elec} can be written as
\begin{equation}
    E_\mathrm{elec}=\sum_{i,j} \mat D_{i,j} \mat H_{0,j,i}.
\end{equation}
Thus, only elements $\{i,j\}$ contribute if $\mat H_{0,j,i}\neq 0$. As long as elements of $\mat D_{j,i}$, where $\mat H_{0,j,i}\neq 0$, are accurately approximated, then the total energy is also estimated accurately, as we have previously demonstrated \cite{Lass_Mohr_Wiebeler_Kühne_Plessl_2018}. This shows that the restriction of the sparsity pattern of $\mat D$ to be the same as for $\hat H_0$, which is implied by the submatrix method, is suitable for electronic structure calculations.

\paragraph{The Non-Orthogonal Local Submatrix Method}
As an extension of the submatrix method, we propose here the NOLSM method, which for the purification in Eq.~\ref{eq:purification} views $\mat D(\mat H_0,\mat S)$ as a matrix function, i.e., the submatrix idea is applied here for the first time simultaneously to two non-orthogonalized matrices.

Accordingly, the steps of the submatrix method are generalized as follows. In the submatrix generation step the sparsity patterns of $\mat H_0$ and $\mat S$ are merged so that a row $i$ is included in the submatrix for column $j$ if either $\mat H_{0,i,j}\neq0$, or $\mat S_{i,j}\neq 0$. Thus, the merged sparsity pattern determines the pair of submatrices $\mathcal T_j(\mat H_0)$ and $\mathcal T_j(\mat S)$ that are equal in size and contain the same indices.
The matrix functions, i.e., the purification
\begin{equation}\label{eq:purification_submatrix}
    \mathcal T_j(\mat D)=\frac{1}{2}\left(\mat I-\sign(\mathcal T_j(\mat S)^{-1}\mathcal T_j(\mat H_0)-\mu \mat I)\right)\mathcal T_j(\mat S)^{-1},
\end{equation}
as well as the computation of the energy-weighted density matrix
\begin{equation}\label{eq:purification_submatrix}
    \mathcal T_j(\mat W)=\mathcal T_j(\mat D)\mathcal T_j(\mat H_0)\mathcal T_j(\mat D)
\end{equation}
are applied directly to the submatrices to yield the submatrix of the density matrix $\mathcal T_j(\mat D)$ and the energy-weighted density matrix $\mathcal T_j(\mat W)$. The entries in columns of these matrices that correspond to the column of the initial matrix form an approximation of the elements in the corresponding columns of the result matrices $\mat D$ and $\mat W$, respectively.

The general arguments for the suitability of the submatrix idea for chemical applications given in Sec.~\ref{sec:submatrix_general} also holds for the non-orthogonal local submatrix method.

\begin{figure*}
    \centering
    \includegraphics[width=0.975\textwidth]{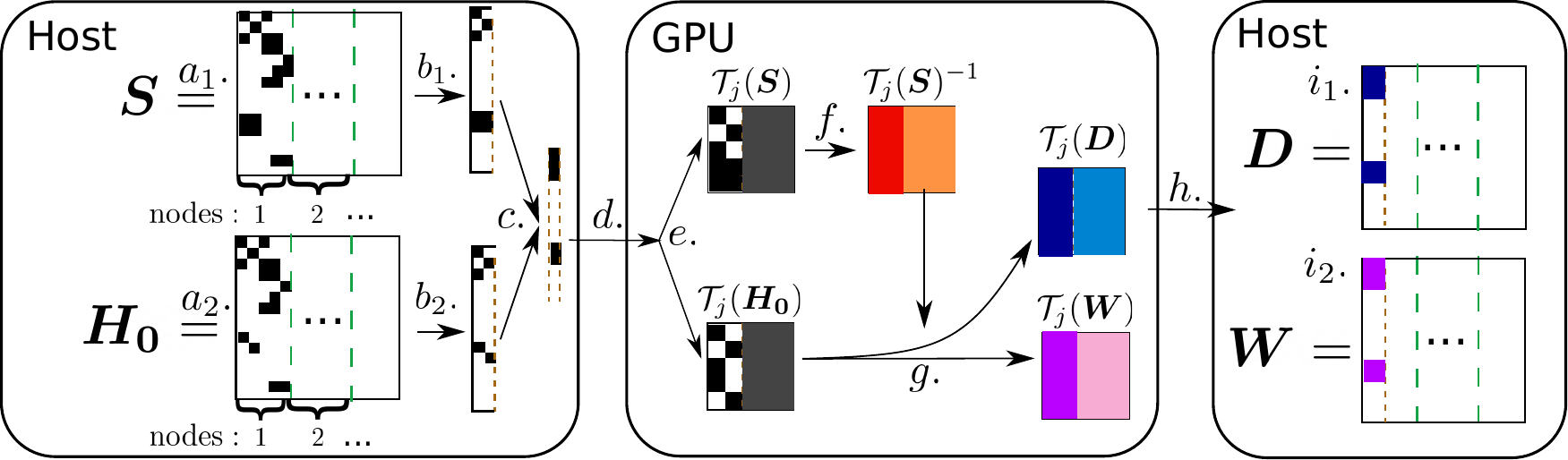}
    \caption{Schematic representation of the steps within our NOLSM method: $a_1.$/$a_2.$ the overlap matrix $\mat S$ ($a_1.$) and the Hamiltonian $\mat H_0$ ($a_2.$) are stored as large sparse matrices, where each column is owned by a node; $b_1.$/$b_2.$ the row indices for the submatrix of a column or multiple columns are extracted; $c.$ row indices of the column from the Hamiltonian and from the overlap matrix are merged; $d.$ row information together with additional data required for the construction of the matrix elements such as atomic positions are transferred to the GPU; $e.$ the matrix elements of the submatrices are generated; $f.$ the submatrix of the overlap matrix is inverted; $g.$ the Hamiltonian submatrix is purified into the density matrix and the energy-weighted density matrix is calculated; $h.$ the result columns are transferred back to the host and inserted at the corresponding places into the sparse matrices of the density matrix $\mat D$ and energy-weighted density matrix $\mat W$ ($i_1.$/$i_2.$).}
    \label{fig:submatrixHS}
\end{figure*}

\paragraph{The Non-Orthogonal Local Submatrix Method with GPUs}
The schematic for the implementation of the NOLSM method with accelerators is shown in Fig.~\ref{fig:submatrixHS}. We discuss here general implementation aspects that are required to obtain an efficient parallel scaling.
Within our implementation, the sparse input matrices $\mat H_0$ and $\mat S$ are generated such that each column is completely owned by one MPI rank (Fig.~\ref{fig:submatrixHS} $a_1.$/$a_2.$). Thus, the row indices required for the submatrices of $\mat H_0$ and $\mat S$ for a column $j$ (Fig.~\ref{fig:submatrixHS} $b_1.$/$b_2.$) can be determined and merged without communicating between the nodes (Fig.~\ref{fig:submatrixHS} $c.$).
The row information and additional data for the construction of the matrix elements are transferred from the host to the GPU (Fig.~\ref{fig:submatrixHS} $d.$). The matrix elements of the submatrices $\mathcal T_j(\mat H_0)$ and $\mathcal T_j(\mat S)$ are not transferred from the host or from other ranks, but are computed locally (Fig.~\ref{fig:submatrixHS} $e.$) on the GPU. The submatrix of the overlap matrix can then be inverted (Fig.~\ref{fig:submatrixHS} $f.$) and the resulting $\mathcal T_j(\mat S)^{-1}$ can be used in the purification (Fig.~\ref{fig:submatrixHS} $g.$) given in Eq.~\ref{eq:purification_submatrix}.
The columns of $\mathcal T_j(\mat D)$ and $\mathcal T_j(\mat W)$ that correspond to the columns of the original matrices hold the approximate matrix elements of the matrices $\mat D$ and $\mat W$. These elements are transferred from the GPUs to the host (Fig.~\ref{fig:submatrixHS} $h.$) and written to the sparse result matrices  $\mat D$ and $\mat W$ (Fig.~\ref{fig:submatrixHS} $i_1.$/$i_2.$), respectively.
This write-back is a local operation so that in total the NOLSM method avoids any inter-node communication during the evaluation of the matrix function.

\color{black}


\subsubsection{Submatrix Combination Heuristics}\label{sec:submatrix_combination}
The NOLSM method as described in section~\ref{sec:localsubmatrix} can be optimized by generating common submatrices for multiple similar columns instead of individual submatrices for each column. However, it is worthwhile to combine more columns: Let $n_i$ and $n_j$ be the dimensions of submatrices $\mathcal T_i$ and $\mathcal T_j$, whereas $n_{i \land j}$ denote the common rows contained both in $\mathcal T_i$ and $\mathcal T_j$. Then a combined or merged submatrix $\mathcal T_{i,j}$ will contain $n_i + n_j - n_{i \land j}$ rows. Considering that the required FLOPs for the evaluation of each submatrix scales cubically, combining $\mathcal T_i$ and $\mathcal T_j$ into $\mathcal T_{i,j}$ yields a speedup if and only if
\begin{equation}\label{eq:combination_criterion}
    {(n_i + n_j - n_{i \land j})}^3 < n_i^3 + n_j^3.
\end{equation}

In this work, we use this relation as a strict acceptance criterion for an iterative combination heuristic. This choice is different from our earlier work that used the spatial location of atoms as guiding properties for the combination of submatrices~\cite{Lass_Schade_Kühne_Plessl_2020}. Also, the presented approach requires no target parameter for the number or size of clusters, but automatically stops when no more improvement of the target metric is found.

For the identification of candidate submatrices $\mathcal T_j$ to be merged into $\mathcal T_i$, the row entries of $\mathcal T_i$ itself are used as a first filter because only submatrices connected in the global sparse matrix tend to have many common row entries (i.e. large $n_{i \land j}$). Given this neighborhood information, valid candidates conforming to the criterion from Eq.~\ref{eq:combination_criterion} are considered iteratively in a sequence that depends on the number of unique elements that $\mathcal T_j$ would add to $\mathcal T_i$, i.e. $n_{j \backslash i} = n_j - n_{i \land j}$. For each iteration $n_{j \backslash i} \in \{0,...,n\}$, the valid merge candidates are first identified and prioritized in parallel and then merged into a common representative $\mathcal T_{i,j}$. This is a similar process to the position update in the k-means clustering employed by Lass et al.~\cite{Lass_Schade_Kühne_Plessl_2020}. However, by using the exact row representation of $\mathcal T_{i,j}$ instead of a spatial position, it allows for the more exact proximity metric (Eq.~\ref{eq:combination_criterion}), while, at the same time, avoiding separate data structures and updates for nearest-neighbor information, thus also facilitating the required scaling to many millions of atoms and submatrices. 
As the combination approach is not performed on individual columns, but on groups of all columns corresponding to each atom, the initial set of submatrices corresponds directly to the number of atoms.
Applying the heuristic from scratch for a system of about 62.5 million atoms takes about two hours on a single compute node and the result can be used for many AIMD steps.  Although in the current work the merging was performed for the full system, parallelization is possible. For that purpose, the system can be subdivided into sufficiently large subcells and the submatrix merging can be performed within every subcell independently. The borders between the subcells can either be treated in a second step by clustering in the border area, while forbidding modifications of clusters that do not touch the border, or simply by neglecting them. In this way, a linear-scaling procedure is obtained.

Yet, instead of fully recomputing the optimal merging of submatrices, if atoms have moved appreciably from the positions for which the merging was originally conducted, simple update strategies of the merging are possible by splitting up merged submatrices and recombining them with other submatrices on-the-fly. However, the development of such update approaches and the necessary row exchanges between the MPI ranks is a topic for future research. 
\color{black}

\subsection{Implementation Innovations}
\subsubsection{Distributed Block Compressed Sparse Row Library: DBCSR}
The Hamiltonian and the overlap matrix have an underlying block-structure originating from the fact that multiple spatial basis functions describe the electronic wave function in the vicinity of an atom and each basis function corresponds to a column of the matrices. The DBCSR sparse matrix library \cite{Borstnik2014}, which handles the sparse matrix operations in CP2K \cite{CP2K}, stores such small blocks in a dense format, while referencing these blocks aka non-zero elements in CSR format. The library is used in this work for the storage and operations on sparse matrices outside of the submatrix method.

\subsubsection{Minimization of Communication}
Due to the favorable parallel properties of the NOLSM method it avoids inter-node communication by construction. The transfers between CPU-main memory and GPUs are minimized by constructing the matrix elements of the submatrices directly on the GPUs. Thus, only metadata, i.e., the row indices required for a submatrix, as well as atom species information, atomic positions for the atoms involved in this submatrix and additional data for the underlying electronic structure method have to be transferred. To reduce the overhead of the matrix-element generation routines on the GPUs, an automatic code generation approach was employed that directly yields expressions for the matrix elements of the overlap matrix between two atomic species that only depend on the distance vector between two atoms. The matrix elements are computed in FP32 to make efficient use of the special function units in NVIDIA GPUs for single-precision floating-point transcendental functions \cite{cudaprogrammingguide}. In addition, we employ a load-balancing between the GPUs in a compute node so that submatrices are assigned to GPUs dynamically.

\subsubsection{Efficient Iterative Matrix Function Solvers on GPU Tensor Cores}
The matrix function for the submatrices, which in the present  case is the purification in Eq.~\ref{eq:purification_submatrix}, can be solved with libraries for dense linear algebra. However, we solve this problem with lower precision linear algebra, specifically making use of the low-precision matrix-multiplications available with tensor cores in GPUs such as the NVIDIA A100 that support FP16 (with FP32 accumulation), bfloat16  (with FP32 accumulation), TensorFloat32 and FP64 \cite{amperewhitepaper}.
For this purpose, the matrix inversion $\mathcal T_i(\mat S)^{-1}$ is performed by an iterative scheme \cite{Higham1997}
\begin{align}
\mat Y_0&=\mathcal T_i(\mat S), \ \ \mat Z_0=\mat I, \ \ \mat V_k=\mat Z_k \mat Y_k \label{eq:inv1}\\
\mat Y_{k+1}&=\frac{1}{2}\mat Y_k(3\mat I-\mat V_k), \ \ \mat Z_{k+1}=\frac{1}{2}(3\mat I-\mat V_k)\mat Z_k  \label{eq:inv2}\\    
\mat Y_0^{-\frac{1}{2}}&=\lim_{k\rightarrow \infty} \mat Z_k, \ \ \mathcal T_i(\mat S)^{-1}=(\mat Y_0^{-\frac{1}{2}})^2\label{eq:inv3}
\end{align}
and the sign function with the Newton-Schulz iteration given in Eq.~\ref{eq:NS1}-\ref{eq:NS2} \cite{Lass_Schade_Kühne_Plessl_2020}. We perform all generalized matrix multiplications (gemm) with the tensor cores in mixed precision: FP16 with FP32-based accumulation. To reduce the overhead, the convergence of Eq.~\ref{eq:NS1}-\ref{eq:NS2} and Eq.~\ref{eq:inv1}-\ref{eq:inv3} is not checked in every iteration, but a fixed number of steps was used.
For the matrix multiplications, kernels from cuBLAS are used. Although individual small- or intermediate-sized matrix multiplications (dimension $\lesssim$ 2000) cannot saturate an NVIDIA A100, we obtain a significant portion of the floating-point performance of the tensor cores by employing three additional strategies: firstly by implementing a suitable caching strategy for the matrices during the iterations in Eq.~\ref{eq:NS1} and \ref{eq:inv2}, secondly by using concurrency via CUDA streams and thirdly by reducing the kernel launch overhead with CUDA graphs~\cite{cudaprogrammingguide}. To reduce the number of required CUDA graphs and the corresponding graph recording overhead, the matrices are padded to multiples of 256. Furthermore, GPU memory allocations are reused.

\section{How Performance Was Measured} 

\subsection{Computational Details}
To investigate the performance of the NOLSM method, we have chosen two kinds of systems: liquid water at ambient condition as a homogeneous system and the HIV-1 capsid solvated in water as an inhomogeneous system.

\subsubsection{Simulation Details}\label{sec:sim_details}
In all of our simulations the GFN-xTB approach is employed in conjunction with a London dispersion correction based on the rational Becke–Johnson damping function \cite{grimme2011effect}. Therein, the electronic states are represented by localized Gaussian-type orbitals. Specifically, every hydrogen atom is represented by two, sulfur by nine, and all other elements used in this work by four basis functions, respectively. As described in Ref.~\cite{CP2K}, we have extended the GFN-xTB method towards periodic boundary conditions. For that purpose, the long-range electrostatic is computed using the fast Fourier transformation (FFT)-based smooth particle mesh Ewald summation with a spline interpolation of fifth order~\cite{doi:10.1063/1.470117}. 

Within our NOLSM method, all submatrices of individual atoms, i.e. all columns corresponding to basis functions of the corresponding nuclei, are coalesced in a single submatrix by default.
Moreover, the chemical potential has been fixed to a value that ensures the overall charge neutrality of the system. To that extent, a bisection-like mechanism as an outer loop for the determination of the chemical potential is implemented to satisfy the charge neutrality constraint. Such a procedure can also be directly integrated into the submatrix method as shown in \cite{Lass_Schade_Kühne_Plessl_2020}. 
In the spirit of the second-generation Car-Parrinello AIMD method, the electronic state is propagated in time by means of fictitious dynamics, thereby completely avoiding the computationally expensive SCF cycle \cite{PhysRevLett.98.066401,WIREs}. Therewith individual AIMD-based dynamical simulated annealing steps with a discretized time-step of 0.5~fs were performed. The modified Langevin-type equation of Eq.~\ref{eq:ModLangevinEq} is integrated using the algorithm of Ricci and Ciccotti \cite{ricci2003algorithms}.

\subsubsection{Water}\label{sec:physical_sys_water}
The water benchmark is derived from the linear-scaling DFT benchmark included with CP2K~\cite{cp2k-lsdft-h2o}. The basic cell contains 32 water molecules and was equilibrated at a temperature of 300 K and a pressure of 1 bar. This cell is cubic with a length of $9.8$~\r{A}. The cell is then repeated in all spatial directions to create a scalable benchmark case. The submatrix combination heuristics is not used for water so that there is one atom per submatrix.

\subsubsection{HIV-1 Capsid}\label{sec:physical_sys_hiv}
\begin{figure}
    \centering
    \includegraphics[width=0.475\textwidth]{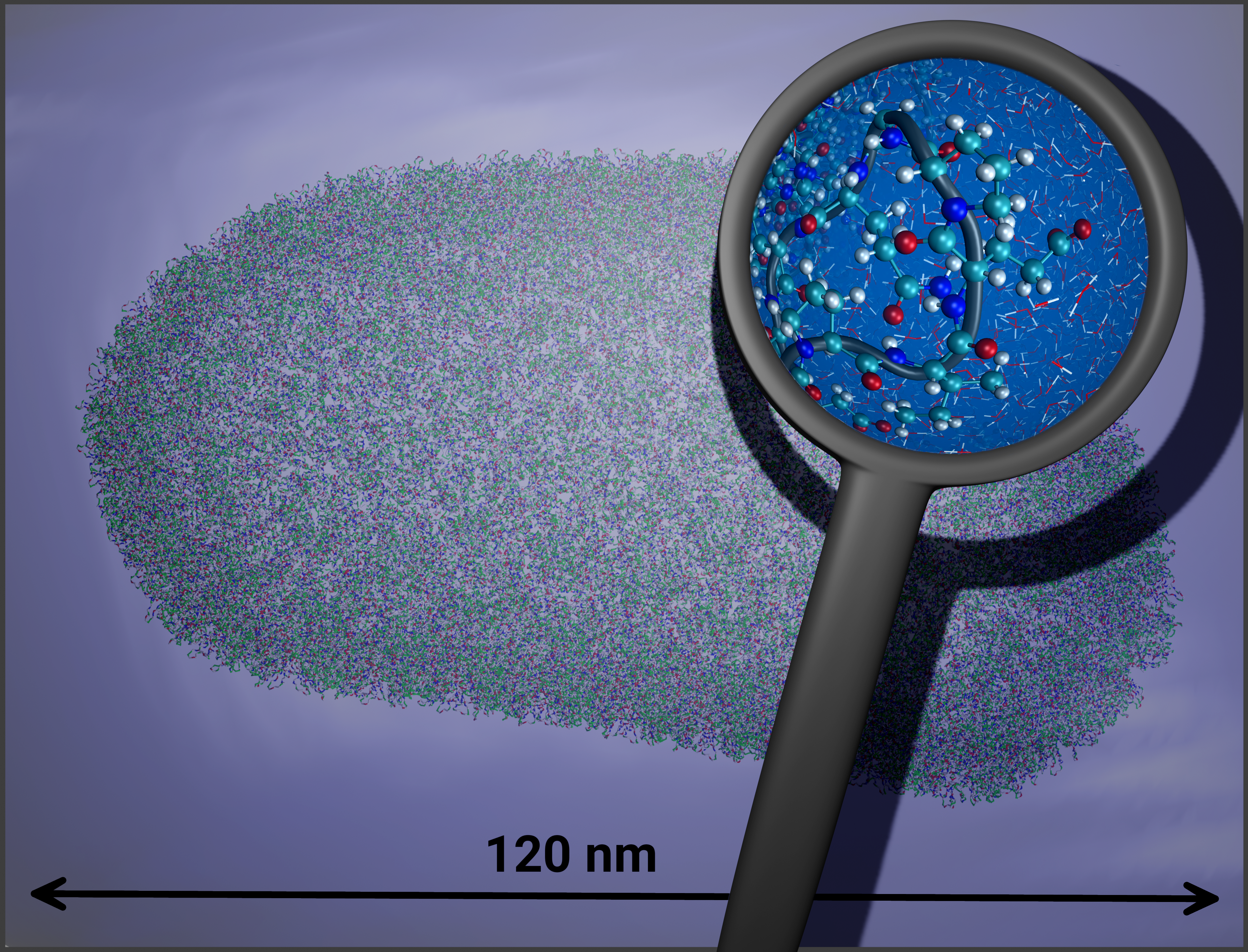}
    \caption{Graphical representation of the present HIV-1 capsid in aqueous solution containing more than 62.5 million atoms.}
    \label{fig:hiv}
\end{figure}

The three-dimensional atomic structure of the entire HIV-1 viral capsid (PDB 3J3Q) was used as a starting point~\cite{zhao2013mature}. The structure is composed of 313,236 amino acid residues with a total of 2,440,800 atoms. Missing hydrogen atoms were added such that all the terminal amino acids, the side chains of lysine, arginine, aspartate, glutamate, and glutamine residues, were all in the charged state. The protonation states of the histidine residues were assigned based on the local hydrogen bonding patterns. 
The capsid was then placed in an orthorhombic unit cell of dimensions 1183.9$\times$800.5$\times$667.8~\r{A} 
and the entire structure was solvated in water. Thereafter, the total charge of the system was neutralized by randomly replacing water molecules 
with sodium ions. The final system, which is shown in Fig.~\ref{fig:hiv} and deposited at \cite{schade_robert_2021_4692508}, has a total of 62,589,576 atoms, i.e. 40,910,985 hydrogen, 1,537,704 carbon, 429,852 nitrogen, 19,689,348 oxygen, 4,059 sodium and 17,628 sulfur atoms, respectively. 

\subsection{Measurements}
\label{sec:measurements}
The main measurements presented here are:

\subsubsection{Wall clock time of the NOLSM method $T_{\mathrm{NOLSM}}$}
The wall clock time $T_{\mathrm{NOLSM}}$ of the core computational routine NOLSM method, i.e., steps $d.$-$h.$ in Fig.~\ref{fig:submatrixHS} are measured. This includes all transfers between host and GPU.

\subsubsection{FLOPs in the NOLSM method $\mathrm{FLOPs}_{\mathrm{NOLSM}}$}
\label{sec:NOLSM_Flops}
The floating-point operations $\mathrm{FLOPs}_{\mathrm{NOLSM}}$ in the FP16/FP32-mixed-precision matrix iterations in the NOLSM method are estimated as $2 n^3$ for a gemm-operation $\mat C=\alpha \mat A\cdot \mat B+\beta \mat C$ with $\mat A,\mat B,\mat C \in \mathbb{R}^{n \times n}$. The construction of the matrix elements of the submatrices, which is performed in FP32, is neglected here because they constitute a small fraction of the total workload and due to the ambiguity of counting the exponential functions as floating-point operations. Other operations that scale as $\mathcal{O}(n^2)$ in the size of the submatrices such as norms and scalings are also neglected in the FLOP count.

\subsubsection{Sustained performance of NOLSM method $P_{\mathrm{NOLSM}}$} To judge the sustained performance obtained from the GPU-acceleration we define the sustained performance $P_{\mathrm{NOLSM}}$ of NOLSM method as $P_{\mathrm{NOLSM}}=\mathrm{FLOPs}_{\mathrm{NOLSM}}/T_{\mathrm{NOLSM}}$.

\subsubsection{Wall time clock for one AIMD step $T_{\mathrm{MD-step}}$}
The wall clock time $T_{\mathrm{MD-step}}$ for a single AIMD step is measured as the average over at least three AIMD steps.
This includes all operations required for one AIMD step, i.e., also IO. This time does not include operations that are only performed once per full MD simulation such as MPI and GPU initialization/finalization or setup of the physical system as well as the heuristic for combining submatrices.

\subsubsection{Time-to-solution $T_\mathrm{sol}$} With the quantities introduced above we define the time-to-solution $T_\mathrm{sol}$ as the wall-clock time $T_{\mathrm{MD-step}}$ for a single AIMD step. This definition is reasonable because the number of AIMD steps per MD calculation can vary greatly depending on the physical or chemical objective of the calculation. Often many thousands or more steps are performed so that operations performed only once per MD calculation can be neglected.


\subsection{HPC System and Environment}
The benchmark calculations have been performed on the JUWELS Booster \cite{juwelsbooster}. The system is ranked as number 7 on the TOP500 list as of winter 2020 with a peak double-precision performance of nearly 71 PFLOP/s~\cite{juwelsboostertop500}. Each of the 936 compute nodes of the JUWELS Booster is a dual-socket system with AMD EPYC 7402 24-core CPUs in NPS4-configuration with 512 GB of DDR4 main memory. Each socket is connected to an individual PCIe-switch that in turn is connected to two NVIDIA A100 GPUs and two Mellanox HDR200 InfiniBand ConnectX6 HCAs. Thus, the theoretical inter-node-communication bandwidth is 800 GBit/s.
The four GPUs per node are fully interconnected with NVLink3.
The cluster interconnect is configured in a DragonFly+ topology with groups of 48 nodes forming a non-blocking cell. The cells are interconnected with 10 links between each cell.

The NVIDIA A100s in the JUWELS BOOSTER have 40 GB of HBM2 memory with a peak memory bandwidth of 1555 GB/s. The theoretical peak performance of the non-tensor-core execution units is
9.7 TFLOP/s in FP64, 19.5 TFLOP/s in FP32 and 78 TFLOP/s in FP16. The peak performances when using the tensor cores are listed as 19.5 TFLOP/s in FP64, 156 TFLOP/s in TF32 and 312 TFLOP/s in FP16 with FP32-based accumulate~\cite{amperewhitepaper}.

The relevant components of the software environment used in this work are GCC 9.3.0, OpenMPI 4.1.0, CUDA NVCC 11.0.221, and CUBLAS 11.0. All benchmarks have been performed with one MPI-rank per node to minimize data replication and 48 CPU-threads per rank. Four CUDA streams are used per GPU and each stream is controlled by a single CPU-thread.

\section{Performance Results}

\subsection{Performance of the NOLSM method}
The performance and scaling of the core steps of the NOLSM method $P_\mathrm{NOLSM}$, i.e., steps $d.$-$h.$ shown in Fig.~\ref{fig:submatrixHS} including transfers to/from GPUs are evaluated in this section. 
\subsubsection{Scaling}
\begin{figure}
    \centering
    \includegraphics[width=0.475\textwidth]{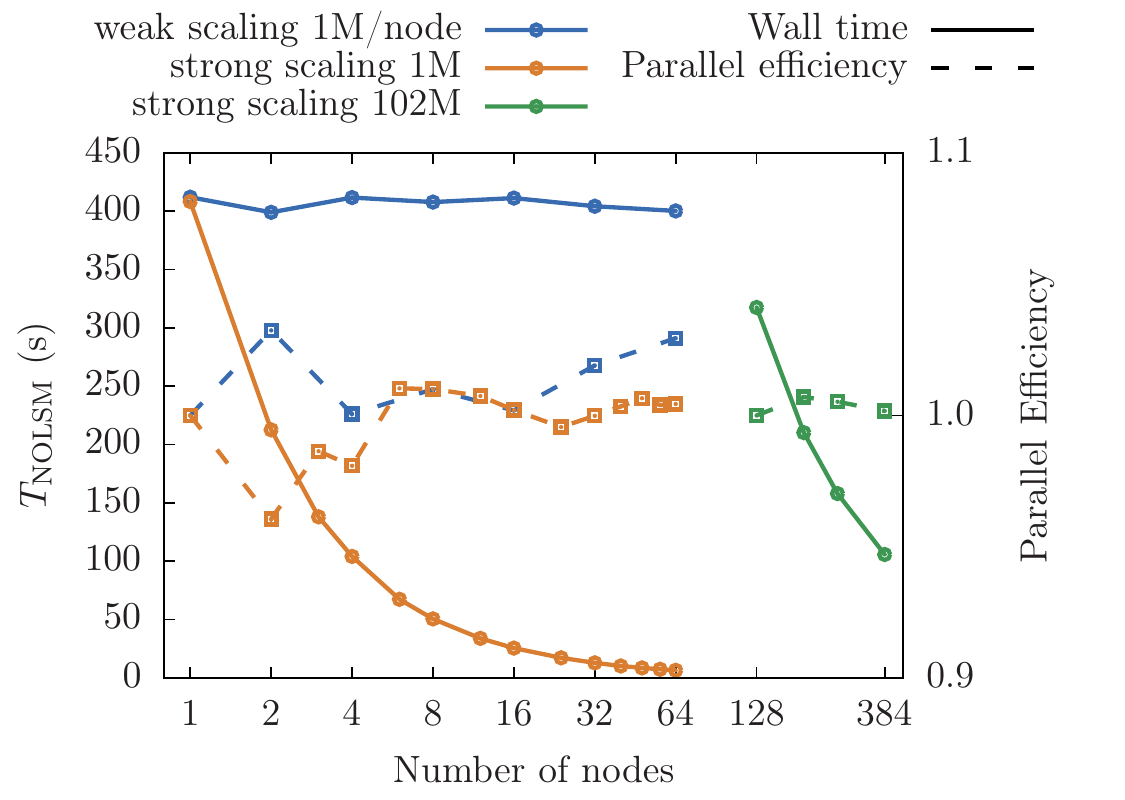}
    \caption{Strong and weak scaling behavior of the NOLSM method for water: weak scaling for 1,022,208 atoms per node (1M) and strong scaling for 1,022,208 atoms (1M) as well as 101,875,968 atoms (102M).}
    \label{fig:NOLSM_T_eff}
\end{figure}

\begin{figure}
    \centering
    \includegraphics[width=0.475\textwidth]{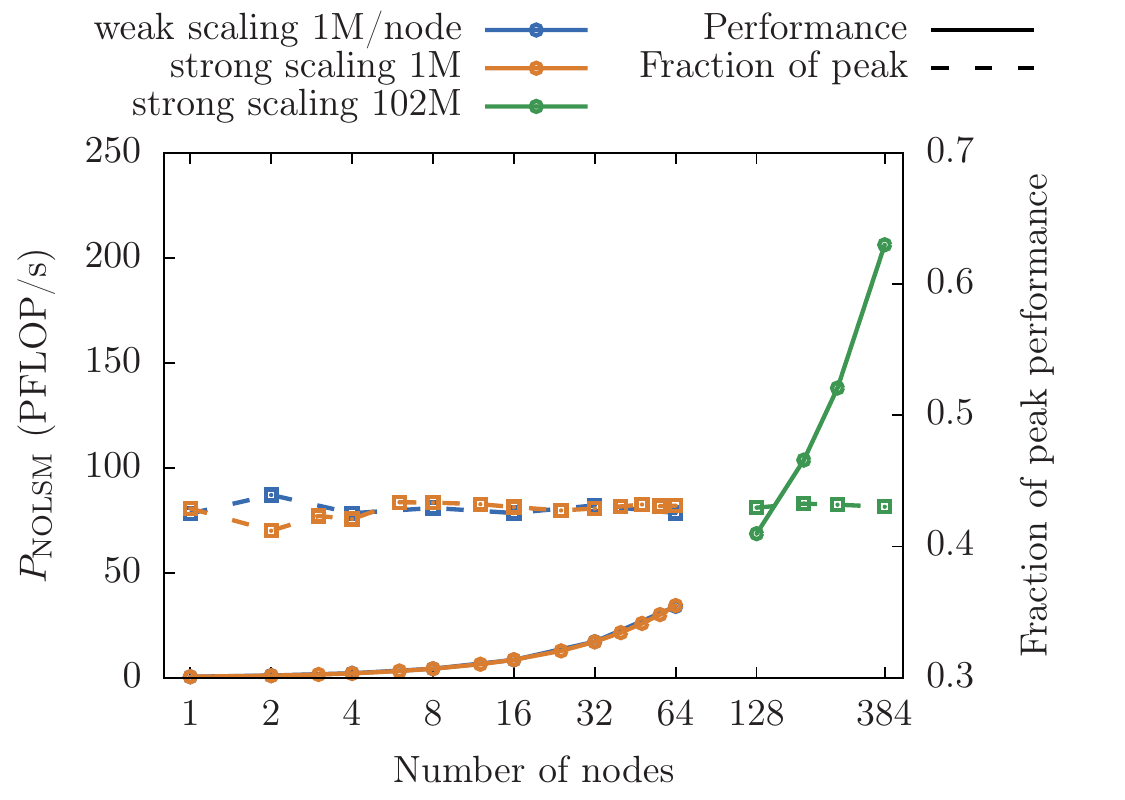}
    \caption{Sustained performance of the NOLSM method and fraction of theoretical peak performance achieved in the strong and weak scaling calculations for water shown in Fig.~\ref{fig:NOLSM_T_eff}.}
    \label{fig:NOLSM_P}
\end{figure}

The physical system used for weak scaling investigations is water, as described in Section~\ref{sec:physical_sys_water}. To allow for scaling from a single node, we have used $22\times22\times22$ basic cells as a basic unit. This basic unit holds 1,022,208 atoms (1M) and is repeated in z-direction. For $n$ compute nodes used, we have $n$ such basic units. For the strong scaling starting from one node, this single basic unit is used. Additionally, strong scaling is also shown for the large-system case with $102\times102\times102$ basic cells corresponding to 101,875,968 atoms (102M). Fig.~\ref{fig:NOLSM_T_eff} shows the wall time and parallel efficiency results. Due to the favorable parallel nature of the non-orthogonalized local submatrix method, a parallel efficiency very close to one is achieved.
In addition, Fig.~\ref{fig:NOLSM_P} shows the achieved floating-point performance $P_\mathrm{NOLSM}$ and fraction of peak performance corresponding to the results shown in 
Fig.~\ref{fig:NOLSM_T_eff}. As defined in Section~\ref{sec:measurements}, the floating-point operations counted here only include matrix multiplications on the tensor cores in mixed precision. Hence, the theoretical peak performance of 312 TFLOP/s per GPU in FP16-based matrix multiplies with FP32 accumulation was used for comparison.
A fraction of about 43 \% of the theoretical peak performance is achieved in this example or, in other words, about 206 PFLOP/s for 384 compute nodes.

\subsubsection{Performance of Matrix Iterations on NVIDIA A100}
\begin{figure}
    \centering
    \includegraphics[width=0.475\textwidth]{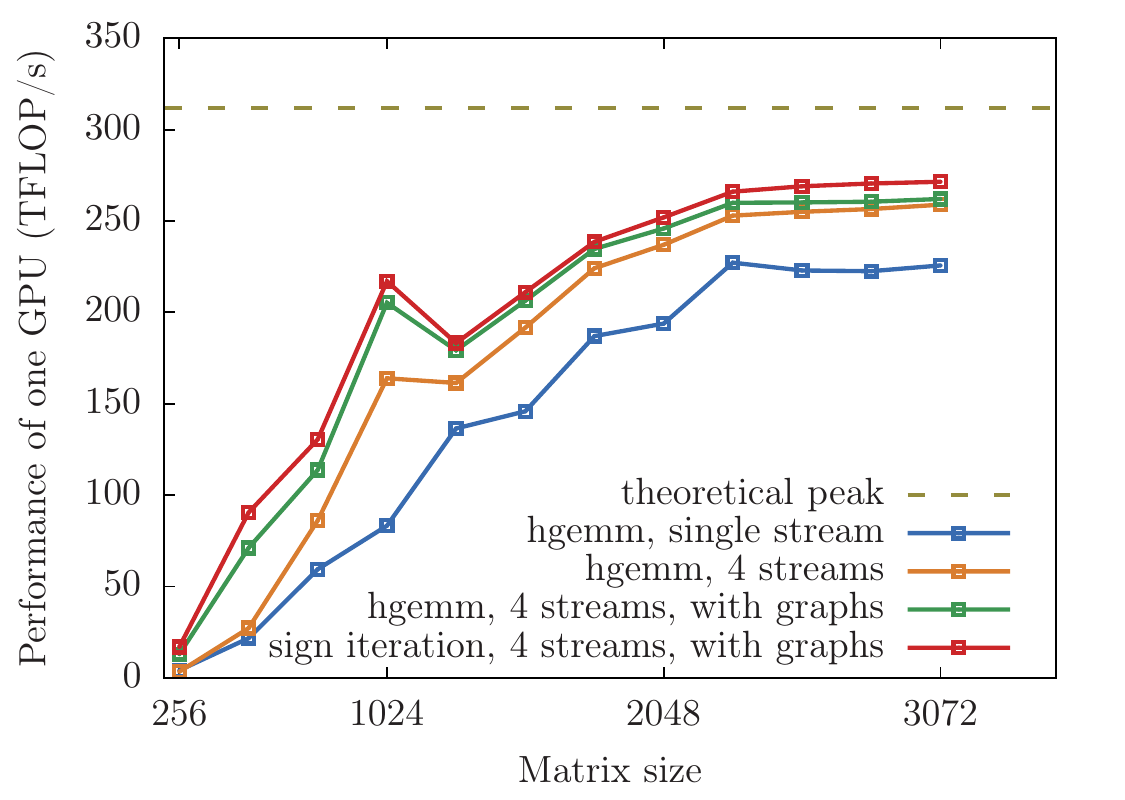}
    \caption{Performance of mixed-precision FP16 with FP32-accumulate matrix multiplications (hgemm) compared to the sign iteration of Eq.~\ref{eq:NS1}-\ref{eq:NS2} with cuBLAS on NVIDIA A100 for different sizes of square matrices and with or without CUDA streams or CUDA graphs.}
    \label{fig:mults}
\end{figure}

Fig.~\ref{fig:mults} shows the increase of performance of cuBLAS-based square matrix multiplications on the tensor cores of the NVIDIA A100 from additional techniques like multiple CUDA streams and CUDA graphs. As a comparison also the performance of the matrix-sign iteration of Eq.~\ref{eq:NS1}-\ref{eq:NS2} while using these techniques is shown. Due to the cache-friendly nature of the matrix iteration, the overall performance is higher than for individual matrix multiplications of the same size. The well-known deficiency of the performance for matrix multiplications of small matrices persists but is mitigated by these additional techniques.

\subsubsection{Effect of Submatrix Combination Heuristics and Transition to HIV-1 System}
\begin{figure}
    \centering
    \includegraphics[width=0.475\textwidth]{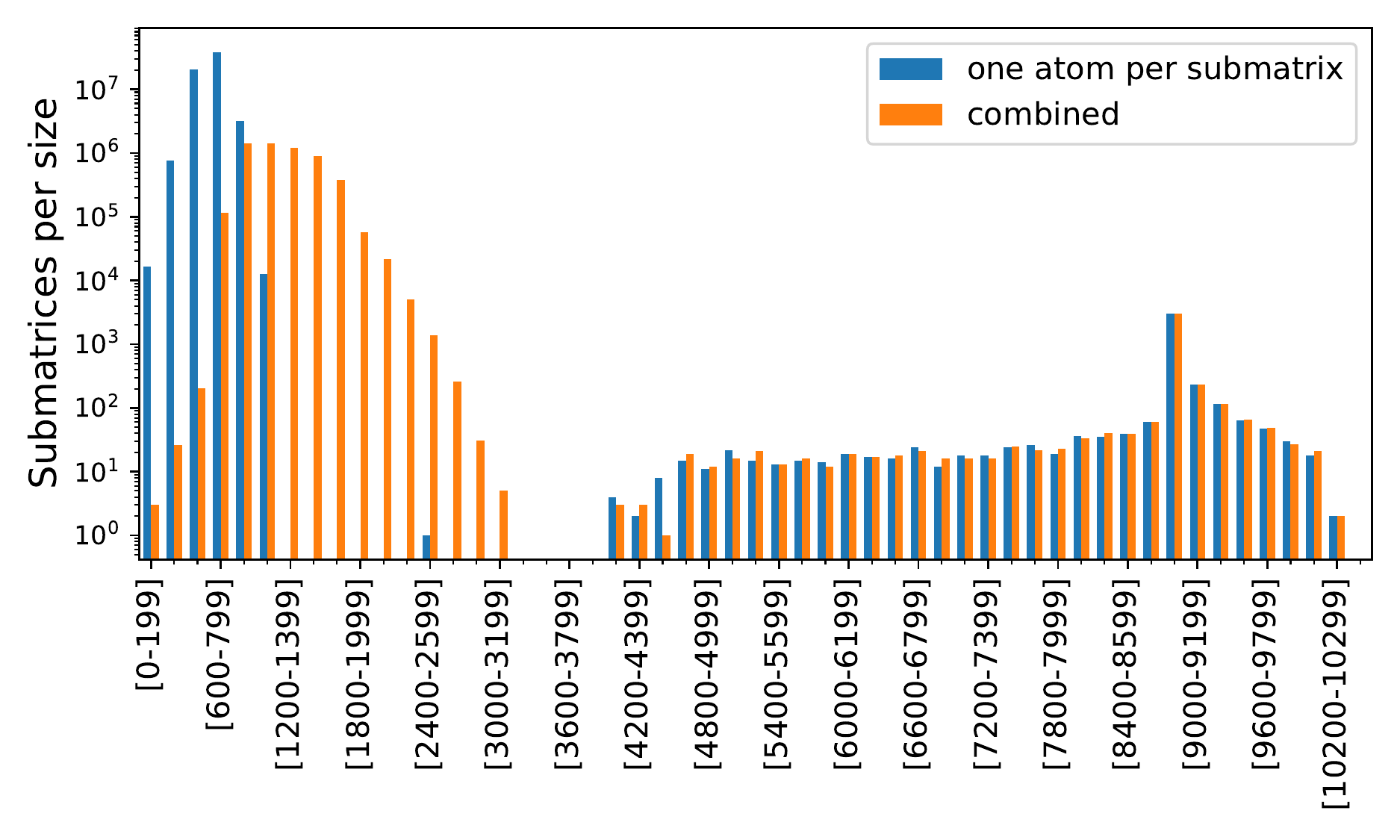}
    \caption{Number of submatrices by groups of sizes $n_i$. The situation with one atom per submatrix resulting in 62,589,576 submatrices is compared to the result of the submatrix combination heuristics that yields 5,536,116 submatrices.}
    \label{fig:submatrixcombination}
\end{figure}

As discussed in Section~\ref{sec:submatrix_combination}, the main benefit of applying the submatrix combination technique is a reduction of the total workload. We investigate the effect with the example of the HIV-1 capsid with about 62.5 million atoms introduced in Section~\ref{sec:physical_sys_hiv}. The heuristic for the combination of submatrices described in Sec.~\ref{sec:submatrix_combination} reduces the cubic work metric that is used as combination criterion in Eq.~\ref{eq:combination_criterion} by a factor of $\approx 1.65$. 
The second benefit of submatrix combination is to increase the average submatrix dimensions into regions where the matrix iterations on GPU reach higher performance as demonstrated in Fig.~\ref{fig:mults}. 
Fig.~\ref{fig:submatrixcombination} shows the effect of submatrix combination on the numbers and sizes of submatrices for the HIV-1 capsid system. We see that the mean and median submatrix sizes increase, most notably by moving the peak of most common sizes from between 400 and 800 elements to between 800 and 1600 elements with an order of magnitude fewer submatrices. Furthermore, we see two disjoint large groups of sizes. The group of smaller submatrices contains up to around 1200 elements before combination and up to 3200 elements afterward, whereas a group of large submatrices exists between 4000 and 10200 elements that changes little by the combination of submatrices. While the combination process itself is completely driven by neighborhood properties encoded in global matrix entries and thus is agnostic of atom types, the large submatrices are formed around sodium atoms. After combination, most of the other submatrices are formed around hydrogen atoms.

The transition from water to the HIV-1 capsid also impacts the calculation of the matrix elements of the submatrices on the GPU. It becomes more elaborate because matrix elements involving species like sulfur require more effort than, for example for hydrogen or oxygen. For example, a matrix element from our automatic code generation approach between two sulfur atoms requires roughly ten times more exponential functions than a matrix element between two oxygen atoms. However, the impact of this is countered by a third effect of the submatrix combination heuristics: not only the cubic target metric of FLOPs during the matrix multiplications is reduced, but at the same time also a quadratic metric of submatrix elements used i.e., $\sum_i n_i^2$ improves by a factor of $\approx 3.3$, which benefits the construction of the submatrices on the GPUs (see step $e.$ in Fig.~\ref{fig:submatrixHS}). 

\begin{figure}
    \centering
    \includegraphics[width=0.475\textwidth]{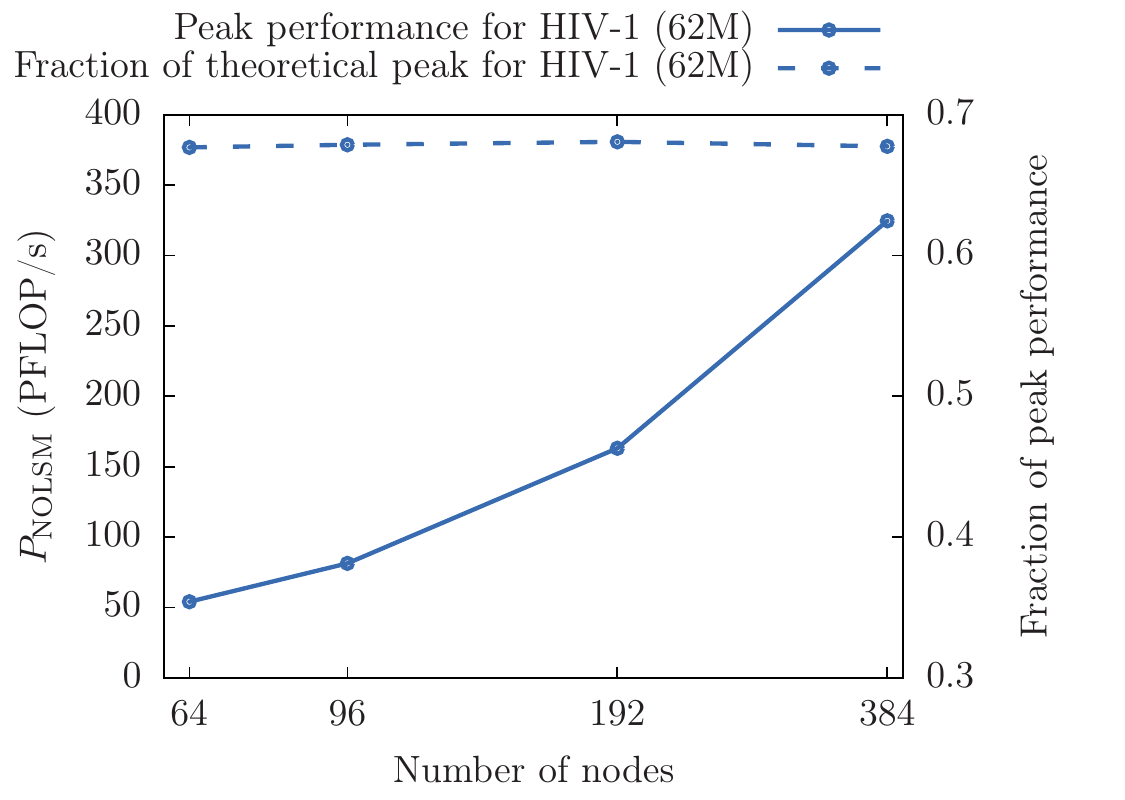}
    \caption{Strong scaling behavior of the NOLSM method for the HIV-1 capsid with $\approx 62$ million atoms.} 
    \label{fig:NOLSM_P_hiv}
\end{figure}

\begin{figure}
    \centering
    \includegraphics[width=0.475\textwidth]{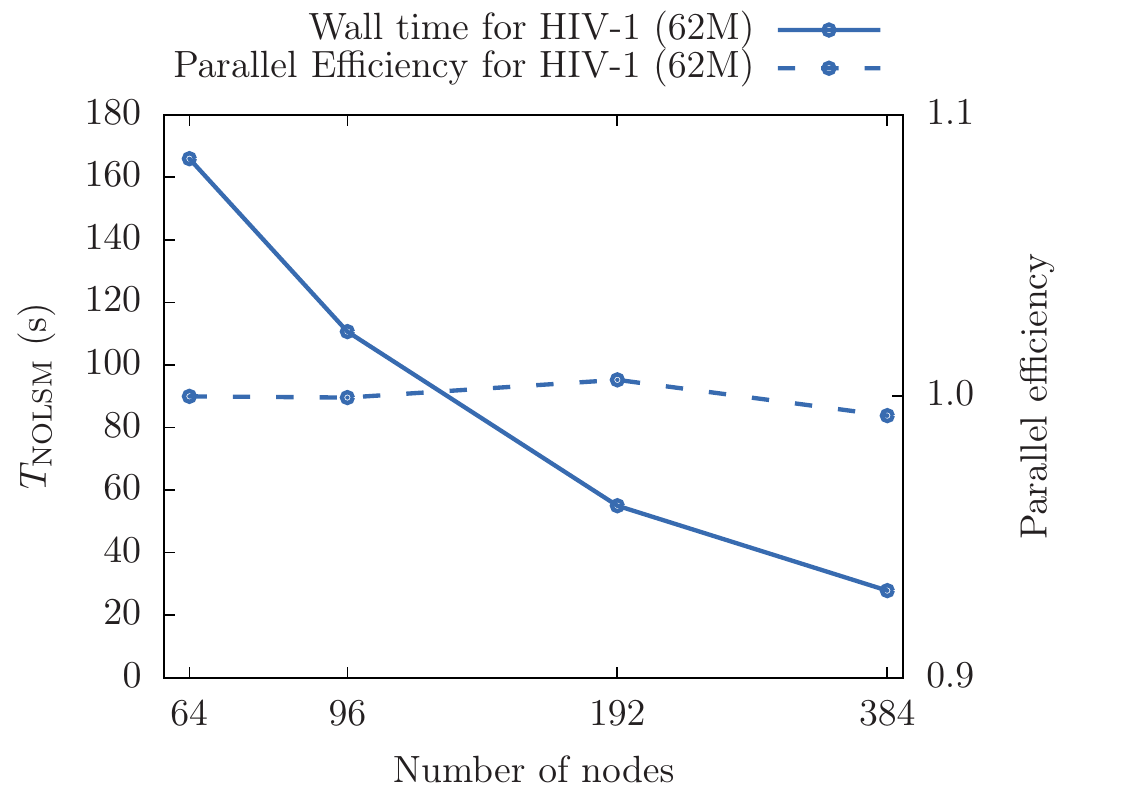}
    \caption{Strong scaling behavior of the NOLSM method for HIV-1 capsid with $\approx 62$ million atoms.} 
    \label{fig:NOLSM_T_hiv}
\end{figure}

Fig.~\ref{fig:NOLSM_P_hiv} shows the resulting strong scaling of the NOLSM method for the sustained performance and fraction of theoretical peak performance after combining submatrices for the HIV-1 capsid.
Please note that the timing for submatrix combination is not included in the measurements because the combination can be precomputed and has to be refreshed only occasionally during an MD calculation.
Thus, the NOLSM method has been shown to reach a sustained performance of about 324.7 PFLOP/s for 384 compute nodes and a fraction of the theoretical peak performance of 67.7 \% for the HIV-1 capsid with about 62.5 million atoms.
In addition, Fig.~\ref{fig:NOLSM_T_hiv} shows the strong-scaling wall time and parallel efficiency for the NOLSM method for the HIV-1 capsid. 


\subsection{Electronic Structure-based Molecular Dynamics}
As shown above, the NOLSM method drastically accelerates the computation of the electronic ground state, which is the core computational routine of an electronic structure-based AIMD calculation. In all previous approaches, this part by far dominated the overall runtime and has thus been a focus of this work. 

\begin{figure}
    \centering
    \includegraphics[width=0.475\textwidth]{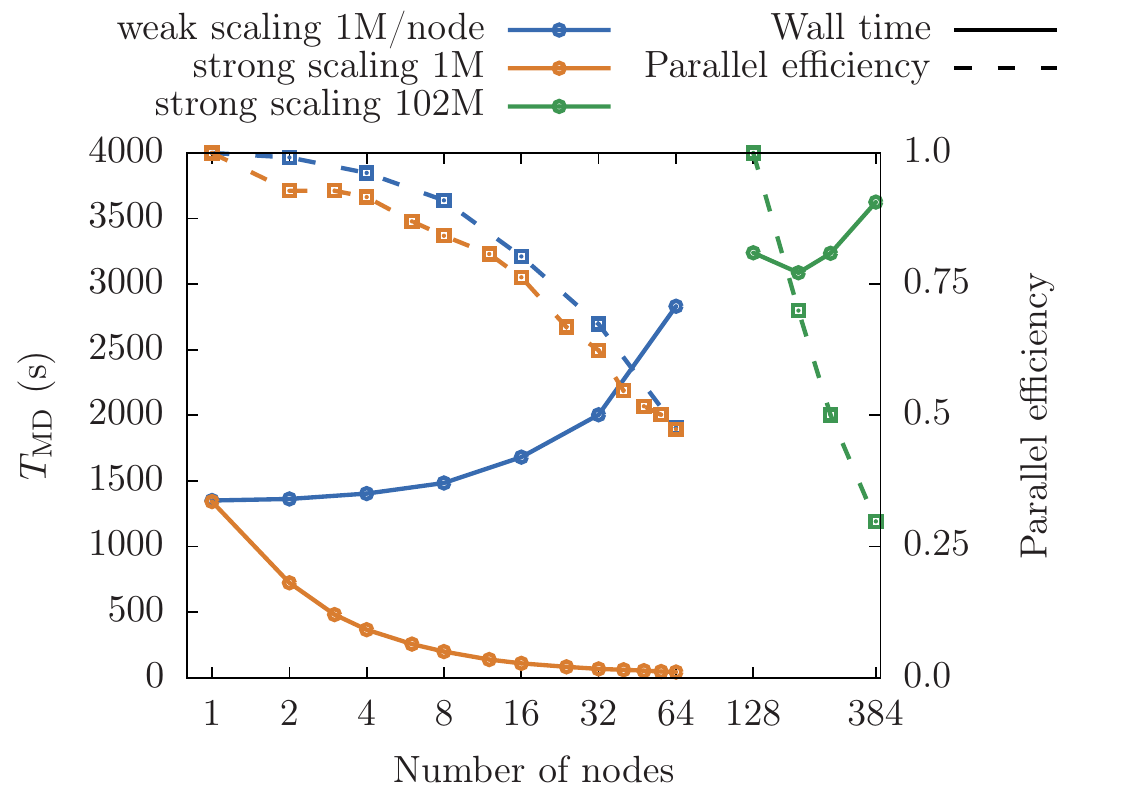}
    \caption{Strong and weak scaling behavior of a full AIMD step for water: weak scaling for 1,022,208 atoms per node (1M) and strong scaling for 1,022,208 atoms (1M) as well as 101,875,968 atoms (102M).}
    \label{fig:MD_T_eff}
\end{figure}

\begin{figure}
    \centering
    \includegraphics[width=0.475\textwidth]{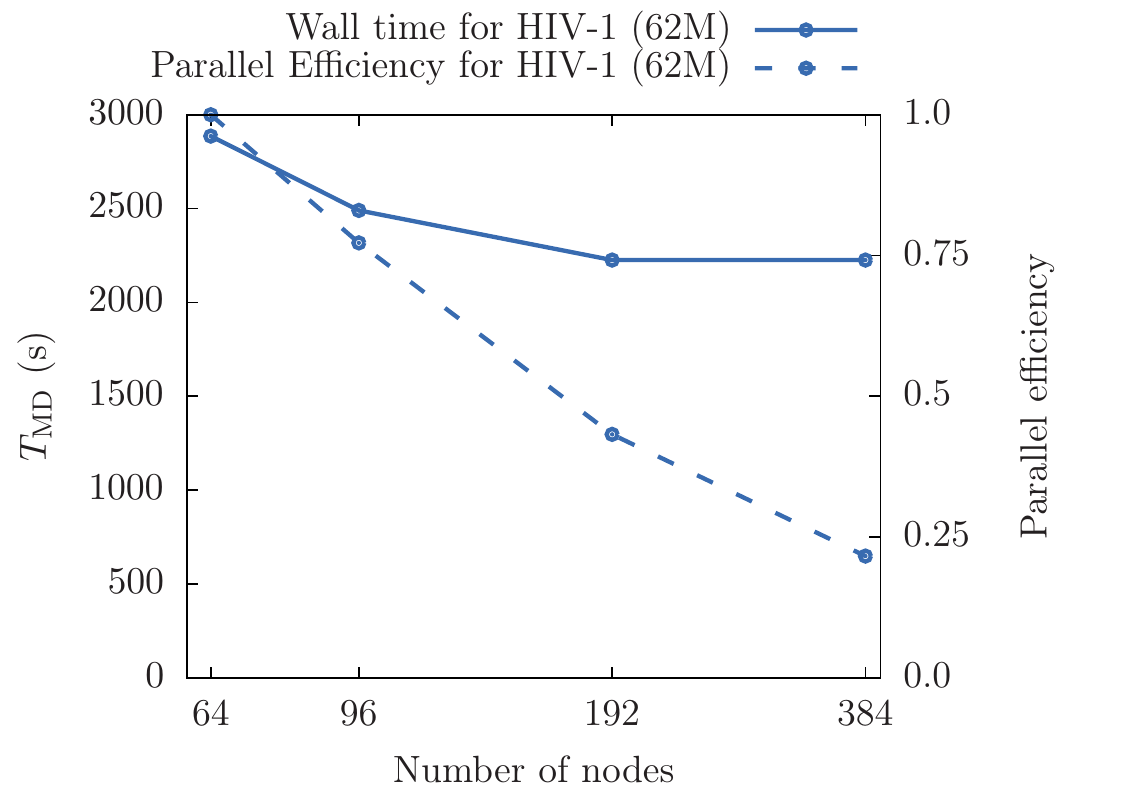}
    \caption{Strong scaling behavior of a full AIMD step for the HIV-1 capsid with $\approx 62$ million atoms.}
    \label{fig:MD_T_eff_hiv}
\end{figure}

Fig.~\ref{fig:MD_T_eff} shows the resulting strong and weak scaling results for water for a complete electronic structure-based AIMD step. The overall scaling of a complete time step suggests that additional effort should be invested in the force evaluation. In total, the use of the NOLSM method has enabled the calculation of an electronic structure-based AIMD step for a system containing more than 100 million atoms in under one hour for the first time.
Fig.~\ref{fig:MD_T_eff_hiv} shows the corresponding results for a full AIMD step on the HIV-1 capsid. The total wall time for a complete time step is well below one hour.

\section{Implications}

The methods developed in this work have far-reaching implications for both scientific applications and extreme-scale simulation methods on modern computer architectures with accelerators.

From the application point of view, our technique allows for performing electronic structure-based simulations for systems with more than 100 million atoms at quantum mechanical accuracy. This contribution allows the application of \textit{ab-inito} simulation methods in life-sciences, where simulation of very large molecules or long time scales are required that are typically only accessible to classical methods. Since our new developments and code improvements will be contributed to the official CP2K code, the broad atomistic-simulation community will be able to directly profit from this work.

On the methodological level, our work can be applied and generalized to a large variety of atomistic simulation and computational science problems. A natural extension is to apply the presented methods to DFT because in CP2K the required construction of the Kohn-Sham Hamiltonian also scales nearly linearly \cite{CP2K}. 
Further, our method to compute exact ensemble averages for observables despite truncation errors in the force computation is neither restricted to AIMD, nor errors introduced by low-precision arithmetic. On the contrary, our approach of using a modified Langevin-type equation is directly applicable to the complete field of MD simulations, in particular for classical methods, and can be used to compensate other forms of approximations that can be modeled as noise, for example, time-step errors, mixed-precision arithmetic or FFT-based Ewald summation for periodic structures \cite{doi:10.1063/1.470117}.

Finally, the submatrix method we have used to compute the matrix sign function in this work is also applicable to other matrix functions, for example, arbitrary polynomials, roots, etc. The precondition is that the sparsity pattern of the initial matrix is approximately preserved under the matrix function. The submatrix method has two main advantages for extreme-scale computing applications: First, it decomposes the problem of computing matrix functions to make it highly parallel while only requiring very little communication. Second, the conversion of the problem from large sparse distributed matrices to much smaller dense local matrices is favorable for GPUs and other accelerators that are optimized for dense matrix algebra. Hence, it is possible to apply the submatrix method in many cases where the core computational problem is a linear eigenproblem that can be reformulated as a matrix function. Examples include the solution of Maxwell's equations in the frequency domain for electrodynamics simulations by casting the problem as a linear eigenproblem \cite{Johnson:01}.

\section*{Acknowledgments}
The authors gratefully acknowledge the Gauss Centre for Supercomputing e.V. (\url{www.gauss-centre.eu}) for funding this project by providing computing time on the GCS Supercomputer JUWELS Booster at Jülich Supercomputing Centre (JSC). Additionally, we would like to thank for funding of this project by computing time provided by the Paderborn Center for Parallel Computing (PC²).
T.D.K. received funding from the European Research Council (ERC) under the European Union’s Horizon 2020 research and innovation program (Grant Agreement No. 716142). T.D.K. and C.P. kindly acknowledge funding from Paderborn University’s research award for ``GreenIT''. Finally, we thank Thomas Müller (JSC), Paul F. Baumeister (JSC), and Markus Hrywniak (NVIDIA) for valuable discussions.

\printbibliography

@Article{MC1953,
  Title                    = {Equation of State Calculations by Fast Computing Machines},
  Author                   = {Metropolis, Nicholas and Rosenbluth, Arianna W. and Rosenbluth, Marshall N. and Teller, Augusta H. and Teller, Edward},
  Journal                  = {J. Chem. Phys.},
  Year                     = {1953},
  Number                   = {6},
  Pages                    = {1087},
  Volume                   = {21},
}

@article{PhysRev.136.A405,
  title = {Correlations in the Motion of Atoms in Liquid Argon},
  author = {Rahman, A.},
  journal = {Phys. Rev.},
  volume = {136},
  number = {2A},
  pages = {A405--A411},
  year = {1964}
}

@article{PhysRevLett.55.2471,
  title = {Unified Approach for Molecular Dynamics and Density-Functional Theory},
  author = {Car, R. and Parrinello, M.},
  journal = {Phys. Rev. Lett.},
  volume = {55},
  number = {22},
  pages = {2471--2474},
  year = {1985}
}

@article{RevModPhys.64.1045,
  title = {Iterative minimization techniques for ab initio total-energy calculations: molecular dynamics and conjugate gradients},
  author = {Payne, M. C. and Teter, M. P. and Allan, D. C. and Arias, T. A. and Joannopoulos, J. D.},
  journal = {Rev. Mod. Phys.},
  volume = {64},
  number = {4},
  pages = {1045--1097},
  year = {1992}
}

@article{PhysRevLett.98.066401,
  title = {Efficient and Accurate Car-Parrinello-like Approach to Born-Oppenheimer Molecular Dynamics},
  author = {K\"uhne, Thomas D. and Krack, Matthias and Mohamed, Fawzi R. and Parrinello, Michele},
  journal = {Phys. Rev. Lett.},
  volume = {98},
  number = {6},
  pages = {066401},
  year = {2007}
}

@article{WaterAIMD,
  title = {Static and Dynamical Properties of Liquid Water from First Principles by a Novel Car-Parrinello-like Approach},
  author = {K\"uhne, Thomas D. and Krack, Matthias and Parrinello, Michele},
  journal = {J. Chem. Theory Comput.},
  volume = {5},
  pages = {235--241},
  year = {2009}
}

@article{Ceriotti,
  title = {Machine learning unifies the modeling of materials and molecules},
  author = {Bartok, Albert P. and others},
  %author = {Bartok, Albert P. and De, Sandip and Poelking, Carl and Bernstein, Noam and Kermode, James R. and Csanyi, Gabor and Ceriotti, Michele},
  journal = {Science Advances},
  volume = {3},
  number = {12},
  pages = {e1701816},
  year = {2017}
}

@article{Tkatchenko,
  title = {Combining Machine Learning and Computational Chemistry for Predictive Insights Into Chemical Systems},
  %author = {Keith, John A. and Vassilev-Galindo, Valentin and Cheng, Bingqing and Chmiela, Stefan and Gastegger, Michael and Müller, Klaus-Robert and Tkatchenko, Alexandre},
  author = {Keith, John A. and others},
  archivePrefix = {arXiv},
  eprint = {2102.06321},
  primaryClass = {physics.chem-ph}
}

@article{UFF,
	title = {UFF, a full periodic table force field for molecular mechanics and molecular dynamics simulations},
	author = {Rappe, A. K. and Casewit, C. J. and Colwell, K. S. and {Goddard III}, W. A. and Skiff, W. M.},
	journal = {J. Am. Chem. Soc.},
	volume = {114},
	number = {25},
	pages = {10024–10035},
	year = {1992}
}

@article{Karplus,
	title = {All-Atom Empirical Potential for Molecular Modeling and Dynamics Studies of Proteins},
	%author = {{MacKerell Jr.}, A. D. and Bashford, D. and Bellott, M. and {Dunbrack Jr.}, R. L. and Evanseck, J. D. and Field, M. J. and Fischer, S. and Gao, J. and Guo, H. and Ha, S. and Joseph-McCarthy, D. and Kuchnir, L. and Kuczera, K. and Lau, F. T. K. and Mattos, C. and Michnick, S. and Ngo, T. and Nguyen, D. T. and Prodhom, B. and Reiher, W. E. and Roux, B. and Schlenkrich, M. and Smith, J. C. and Stote, R. and Straub, J. and Watanabe, M. and Wiorkiewicz-Kuczera, J. and Yin, D. and Karplus, M.},
	author = {{MacKerell Jr.}, A. D. and others}, 
	journal = {J. Phys. Chem. B},
	volume = {102},
	number = {18},
	pages = {3586–3616},
    year = {1998}
}

@article{mcweeny1960some,
  title={Some recent advances in density matrix theory},
  author={McWeeny, Rev},
  journal={Reviews of Modern Physics},
  volume={32},
  number={2},
  pages={335},
  year={1960},
  publisher={APS}
}

@article{WIREs,
  title = {Second generation Car–Parrinello molecular dynamics},
  author = {K\"uhne, Thomas D.},
  journal = {WIREs Comput. Mol. Sci.},
  volume = {4},
  pages = {391-406},
  year = {2014}
}

@article{kuhne2018disordered,
  title={Disordered crystals from first principles I: Quantifying the configuration space},
  author={K{\"u}hne, Thomas D and Prodan, Emil},
  journal={Annals of Physics},
  volume={391},
  pages={120--149},
  year={2018},
  publisher={Elsevier}
}

@article{kuhne2020disordered,
  title={Disordered crystals from first principles II: Transport coefficients},
  author={K{\"u}hne, Thomas D and Heske, Julian and Prodan, Emil},
  journal={Annals of Physics},
  volume={421},
  pages={168290},
  year={2020},
  publisher={Elsevier}
}

@article{SM,
  title={Linear scaling density matrix search based on sign matrices},
  author={Nemeth, Karoly and Scuseria, Gustavo E.},
  journal={J. Chem. Phys.},
  volume={113},
  pages={6035--6041},
  year={2000}
}

@article{ProdanKohn,
  title = {Nearsightedness of electronic matter},
  author = {Prodan, E. and Kohn, W.},
  journal = {Proc. Nat. Acad. Sci. USA},
  volume = {102},
  number = {33},
  pages = {11635-11638},
  year = {2005}
}

@article{RevModPhys.71.1085,
  title = {Linear scaling electronic structure methods},
  author = {Goedecker, Stefan},
  journal = {Rev. Mod. Phys.},
  volume = {71},
  number = {4},
  pages = {1085--1123},
  year = {1999}
}

@article{PhysRevLett.66.1438,
  title = {Direct calculation of electron density in density-functional theory},
  author = {Yang, Weitao},
  journal = {Phys. Rev. Lett.},
  volume = {66},
  number = {11},
  pages = {1438--1441},
  year = {1991}
}

@article{PhysRevLett.69.3547,
  title = {Large scale electronic structure calculations},
  author = {Galli, Giulia and Parrinello, Michele},
  journal = {Phys. Rev. Lett.},
  volume = {69},
  number = {24},
  pages = {3547--3550},
  year = {1992}
}

@article{Richters,
  title = {Self-consistent field theory based molecular dynamics with linear system-size scaling},
  author = {Richters, Dorothee and K\"uhne, Thomas D.},
  journal = {J. Chem. Phys.},
  volume = {140},
  number = {13},
  pages = {134109},
  year = {2014}
}

@article{Computation,
  title = {Accurate Sampling with Noisy Forces from Approximate Computing},
  author = {Rengaraj, Varadarajan and Lass, Michael and Plessl, Christian and K\"uhne, Thomas D.},
  journal = {Computation},
  volume = {8},
  number = {39},
  pages = {1-11},
  year = {2020}
}

@article{AC1,
  title = {Changing Computing Paradigms Towards Power Efficiency},
  author = {Klavik, P. and Malossi, A.C.I. and Bekas, C. and Curioni, A.},
  journal = {Philos. Trans. R. Soc. A Math. Phys. Eng. Sci.},
  volume = {39},
  number = {2018},
  pages = {372},
  year = {2014}
}

@article{AC2,
  title = {Approximate Computing},
  author = {Plessl, C. and Platzner, M. and Schreier, P. J.},
  journal = {Inform. Spektrum},
  volume = {15},
  pages = {396–399},
  year = {2015}
}

@article{sign-method,
  title = {Linear Scaling Self-Consistent Field Calculations with Millions of Atoms in the Condensed Phase},
  author = {VandeVondele, J. and  Borstnik, U. and Hutter, J.},
  journal = {J. Chem. Theory Comput.},
  volume = {8},
  number = {10},
  pages = {3565–3573},
  year = {2012}
}

@article{Richters_Lass_Walther_Plessl_Kühne_2019, title={A General Algorithm to Calculate the Inverse Principal p-th Root of Symmetric Positive Definite Matrices}, volume={25}, 
%DOI={10.4208/cicp.OA-2018-0053}, 
number={2}, journal={Communications in Computational Physics}, publisher={Global Science Press}, author={Richters, Dorothee and Lass, Michael and Walther, Andrea and Plessl, Christian and Kühne, Thomas}, year={2019}, pages={564–585} }

@article{CP2K,
  title = {{CP2K}: An electronic structure and molecular dynamics software package - Quickstep: Efficient and accurate electronic structure calculations},
  %author = {Kühne, {Thomas D.} and Iannuzzi, Marcella and {Del Ben}, Mauro and Rybkin, {Vladimir V.} and Seewald, Patrick and Stein, Frederick and Laino, Teodoro and Khaliullin, {Rustam Z.} and Schütt, Ole and Schiffmann, Florian and Golze, Dorothea and Wilhelm, Jan and Chulkov, Sergey and Bani-Hashemian, {Mohammad Hossein} and Weber, Valery and Borstnik,  Urban and Taillefumier, Mathieu and Jakobovits, {Alice Shoshana} and Lazzaro, Alfio and Pabst, Hans and Müller, Tiziano and Schade, Robert and Guidon, Manuel and Andermatt, Samuel and Holmberg, Nico and Schenter, {Gregory K.} and Hehn, Anna and Bussy, Augustin and Belleflamme, Fabian and Tabacchi, Gloria and Glö{\ss}, Andreas and Lass, Michael and Bethune, Iain and Mundy, Christopher J. and Plessl, Christian and Watkins, Matt and VandeVondele, Joost and Krack, Matthias and Hutter, Jürg},
  author = {Kühne, {Thomas D.} and others},
  journal = {J. Chem. Phys.},
  volume = {152},
  number = {19},
  pages = {194103},
  year = {2020}
}

@inproceedings{Lass_Mohr_Wiebeler_Kühne_Plessl_2018, place={New York, NY, USA}, title={A Massively Parallel Algorithm for the Approximate Calculation of Inverse p-th Roots of Large Sparse Matrices}, %DOI={10.1145/3218176.3218231}, 
booktitle={Proc. Platform for Advanced Scientific Computing (PASC) Conference}, publisher={ACM}, author={Lass, Michael and Mohr, Stephan and Wiebeler, Hendrik and Kühne, Thomas and Plessl, Christian}, year={2018} }

@inproceedings{Lass_Schade_Kühne_Plessl_2020, place={Los Alamitos, CA, USA}, title={A Submatrix-Based Method for Approximate Matrix Function Evaluation in the Quantum Chemistry Code {CP2K}}, 
%DOI={10.1109/SC41405.2020.00084}, 
booktitle={Proc. International Conference for High Performance Computing, Networking, Storage and Analysis (SC)}, publisher={IEEE Computer Society}, author={Lass, Michael and Schade, Robert and Kühne, Thomas and Plessl, Christian}, year={2020}, pages={1127–1140} }

@inproceedings{gygi2006large,
  title={Large-scale electronic structure calculations of high-Z metals on the {BlueGene/L} platform},
  %author={Gygi, Francois and Draeger, Erik W and Schulz, Martin and De Supinski, Bronis R and Gunnels, John A and Austel, Vernon and Sexton, James C and Franchetti, Franz and Kral, Stefan and Ueberhuber, Christoph W and others},
  author={Gygi, Francois and others},
  booktitle={Proceedings of the 2006 ACM/IEEE conference on Supercomputing},
  pages={45},
  year={2006}
}

@article{gygi2008architecture,
  title={Architecture of {Qbox}: A scalable first-principles molecular dynamics code},
  author={Gygi, Francois},
  journal={IBM Journal of Research and Development},
  volume={52},
  number={1.2},
  pages={137--144},
  year={2008},
  publisher={IBM}
}

@article{hutter2005dual,
  title={Dual-level parallelism for ab initio molecular dynamics: Reaching teraflop performance with the {CPMD} code},
  author={Hutter, J{\"u}rg and Curioni, Alessandro},
  journal={Parallel Computing},
  volume={31},
  number={1},
  pages={1--17},
  year={2005},
  publisher={Elsevier}
}

@article{hutter2005car,
  title={Car--Parrinello molecular dynamics on massively parallel computers},
  author={Hutter, J{\"u}rg and Curioni, Alessandro},
  journal={ChemPhysChem},
  volume={6},
  number={9},
  pages={1788--1793},
  year={2005},
  publisher={Wiley Online Library}
}

@inproceedings{das2019fast,
  title={Fast, scalable and accurate finite-element based ab initio calculations using mixed precision computing: 46 {PFLOPS} simulation of a metallic dislocation system},
  author={Das, Sambit and Motamarri, Phani and Gavini, Vikram and Turcksin, Bruno and Li, Ying Wai and Leback, Brent},
  booktitle={Proceedings of the International Conference for High Performance Computing, Networking, Storage and Analysis},
  pages={1--11},
  year={2019}
}

@article{motamarri2020dft,
  title={{DFT}-{FE}--A massively parallel adaptive finite-element code for large-scale density functional theory calculations},
  %author={Motamarri, Phani and Das, Sambit and Rudraraju, Shiva and Ghosh, Krishnendu and Davydov, Denis and Gavini, Vikram},
  author={Motamarri, Phani and others},
  journal={Computer Physics Communications},
  volume={246},
  pages={106853},
  year={2020},
  publisher={Elsevier}
}

@misc{cp2k-lsdft-h2o,
url="https://github.com/cp2k/cp2k/blob/028e7b8381f1bc85b52fb82ab205a43ab6f0c339/benchmarks/QS_DM_LS/H2O-dft-ls.inp"
}

@manual{cudaprogrammingguide,
 title        = {{CUDA} {C++} Programming Guide},
  author       = {{NVIDIA Corporation}}, 
  organization = {NVIDIA Corporation},
  year         = 2021,
  url="https://docs.nvidia.com/cuda/pdf/CUDA_C_Programming_Guide.pdf"
}

@misc{amperewhitepaper,
url="https://www.nvidia.com/content/dam/en-zz/Solutions/Data-Center/nvidia-ampere-architecture-whitepaper.pdf"
}

@Article{doi:10.1002/zamm.19330130111,
  author  = {Schulz, Günther},
   journal = {ZAMM - Journal of Applied Mathematics and Mechanics / Zeitschrift für Angewandte Mathematik und Mechanik},
   title   = {{Iterative Berechung der reziproken Matrix}},
   year    = 1933,
   number  = 1,
   pages   = {57-59},
   volume  = 13,
   %doi     = {10.1002/zamm.19330130111},
   eprint-xx  = {https://onlinelibrary.wiley.com/doi/pdf/10.1002/zamm.19330130111},
   url-xx     = {https://onlinelibrary.wiley.com/doi/abs/10.1002/zamm.19330130111},
}

@Article{Higham1997,
  author    = {Nicholas J. Higham},
  journal   = {Numerical Algorithms},
  title     = {Stable iterations for the matrix square root},
  year      = 1997,
  number    = 2,
  pages     = {227--242},
  volume    = 15,
  %doi       = {10.1023/a:1019150005407},
  publisher = {Springer},
}

@Article{Borstnik2014,
  author    = {Urban Bor{\v{s}}tnik and Joost VandeVondele and Val{\'{e}}ry Weber and Jürg Hutter},
  journal   = {Parallel Computing},
  title     = {Sparse matrix multiplication: {The} distributed block-compressed sparse row library},
  year      = 2014,
  month     = may,
  number    = {5-6},
  pages     = {47--58},
  volume    = 40,
  %doi       = {10.1016/j.parco.2014.03.012},
  groups    = {CP2K},
  publisher = {Elsevier},
}

@mics{juwelsbooster,
title="Hardware Configuration of the JUWELS Booster Module",
url="https://www.fz-juelich.de/ias/jsc/EN/Expertise/Supercomputers/JUWELS/Configuration/Configuration_node.html"}

@mics{juwelsboostertop500,
title="{JUWELS} Booster TOP 500 entry",
url="https://www.top500.org/system/179894/"
}

@article{nakata2020large,
  title={Large scale and linear scaling {DFT} with the {CONQUEST} code},
  author={Nakata, Ayako and others},
  journal={The Journal of chemical physics},
  volume={152},
  number={16},
  pages={164112},
  year={2020},
  publisher={AIP Publishing LLC}
}

@article{hasegawa2014performance,
  title={Performance evaluation of ultra-large-scale first-principles electronic structure calculation code on the K computer},
  author={Hasegawa, Yukihiro and others},
  journal={The International journal of high performance computing applications},
  volume={28},
  number={3},
  pages={335--355},
  year={2014},
  publisher={Sage Publications Sage UK: London, England}
}

@inproceedings{zhao2009linearly,
  title={The linearly scaling 3D fragment method for large scale electronic structure calculations},
  author={Zhao, Zhengji and others},
  %author={Zhao, Zhengji and Meza, Juan and Lee, Byounghak and Shan, Hongzhang and Strohmaier, Erich and Bailey, David and Wang, Lin-Wang},
  booktitle={Journal of Physics: Conference Series},
  volume={180},
  number={1},
  pages={012079},
  year={2009},
  organization={IOP Publishing}
}

@article{zhao2013mature,
  title={Mature {HIV-1} capsid structure by cryo-electron microscopy and all-atom molecular dynamics},
  author={Zhao, Gongpu and others},
  journal={Nature},
  volume={497},
  number={7451},
  pages={643--646},
  year={2013},
  publisher={Nature Publishing Group}
}

@article{andermatt2016combining,
  title={Combining linear-scaling {DFT} with subsystem {DFT} in Born--Oppenheimer and Ehrenfest molecular dynamics Simulations: from molecules to a virus in solution},
  author={Andermatt, Samuel and Cha, Jinwoong and Schiffmann, Florian and VandeVondele, Joost},
  journal={Journal of chemical theory and computation},
  volume={12},
  number={7},
  pages={3214--3227},
  year={2016},
  publisher={ACS Publications}
}

@article{grimme2011effect,
  title={Effect of the damping function in dispersion corrected density functional theory},
  author={Grimme, Stefan and Ehrlich, Stephan and Goerigk, Lars},
  journal={Journal of computational chemistry},
  volume={32},
  number={7},
  pages={1456--1465},
  year={2011},
  publisher={Wiley Online Library}
}

@article{karhan2014role,
  title={On the role of interfacial hydrogen bonds in “on-water” catalysis},
  author={Karhan, Kristof and Khaliullin, Rustam Z and K{\"u}hne, Thomas D},
  journal={The Journal of chemical physics},
  volume={141},
  number={22},
  pages={12B632\_1},
  year={2014},
  publisher={American Institute of Physics}
}

@inproceedings{jain2016openatom,
  title={Openatom: Scalable ab-initio molecular dynamics with diverse capabilities},
  %author={Jain, Nikhil and Bohm, Eric and Mikida, Eric and Mandal, Subhasish and Kim, Minjung and Jindal, Prateek and Li, Qi and Ismail-Beigi, Sohrab and Martyna, Glenn J and Kale, Laxmikant V},
  author={Jain, Nikhil and others},
  booktitle={International Conference on High Performance Computing},
  pages={139--158},
  year={2016},
  organization={Springer}
}

@article{prentice2020onetep,
  title={The {ONETEP} linear-scaling density functional theory program},
  author={Prentice, Joseph CA and others},
  journal={The Journal of chemical physics},
  volume={152},
  number={17},
  pages={174111},
  year={2020},
  publisher={AIP Publishing LLC}
}

@article{wilkinson2014hybrid,
  title={Hybrid {MPI}-{OpenMP} parallelism in the {ONETEP} linear-scaling electronic structure code: Application to the delamination of cellulose nanofibrils},
  author={Wilkinson, Karl A and Hine, Nicholas DM and Skylaris, Chris-Kriton},
  journal={Journal of chemical theory and computation},
  volume={10},
  number={11},
  pages={4782--4794},
  year={2014},
  publisher={ACS Publications}
}

@article{vandevondele2012linear,
  title={Linear scaling self-consistent field calculations with millions of atoms in the condensed phase},
  author={VandeVondele, Joost and Borstnik, Urban and Hutter, Jurg},
  journal={Journal of chemical theory and computation},
  volume={8},
  number={10},
  pages={3565--3573},
  year={2012},
  publisher={ACS Publications}
}

@inproceedings{fattebert2016modeling,
  title={Modeling dilute solutions using first-principles molecular dynamics: computing more than a million atoms with over a million cores},
  author={Fattebert, Jean-Luc and Osei-Kuffuor, Daniel and Draeger, Erik W and Ogitsu, Tadashi and Krauss, William D},
  booktitle={SC'16: Proceedings of the International Conference for High Performance Computing, Networking, Storage and Analysis},
  pages={12--22},
  year={2016},
  organization={IEEE}
}

@inproceedings{nomura2014metascalable,
  title={Metascalable quantum molecular dynamics simulations of hydrogen-on-demand},
  %author={Nomura, Ken-ichi and Kalia, Rajiv K and Nakano, Aiichiro and Vashishta, Priya and Shimamura, Kohei and Shimojo, Fuyuki and Kunaseth, Manaschai and Messina, Paul C and Romero, Nichols A},
  author={Nomura, Ken-ichi and others},
  booktitle={SC'14: Proceedings of the International Conference for High Performance Computing, Networking, Storage and Analysis},
  pages={661--673},
  year={2014},
  organization={IEEE}
}

@article{Johnson:01,
author = {Steven G. Johnson and J. D. Joannopoulos},
journal = {Opt. Express},
keywords = {Numerical approximation and analysis; Birefringence; Fourier transforms; Narrow band filters; Photonic crystals; Signal processing; Spatial resolution},
number = {3},
pages = {173--190},
publisher = {OSA},
title = {Block-iterative frequency-domain methods for Maxwell's equations in a planewave basis},
volume = {8},
month = {Jan},
year = {2001},
%doi = {10.1364/OE.8.000173},
abstract = {We describe a fully-vectorial, three-dimensional algorithm to compute the definite-frequency eigenstates of Maxwell's equations in arbitrary periodic dielectric structures, including systems with anisotropy (birefringence) or magnetic materials, using preconditioned block-iterative eigensolvers in a planewave basis. Favorable scaling with the system size and the number of computed bands is exhibited. We propose a new effective dielectric tensor for anisotropic structures, and demonstrate that O($\Delta$x2) convergence can be achieved even in systems with sharp material discontinuities. We show how it is possible to solve for interior eigenvalues, such as localized defect modes, without computing the many underlying eigenstates. Preconditioned conjugate-gradient Rayleigh-quotient minimization is compared with the Davidson method for eigensolution, and a number of iteration variants and preconditioners are characterized. Our implementation is freely available on the Web.},
}

@article{doi:10.1021/acs.jctc.7b00118,
author = {Grimme, Stefan and Bannwarth, Christoph and Shushkov, Philip},
title = {A Robust and Accurate Tight-Binding Quantum Chemical Method for Structures, Vibrational Frequencies, and Noncovalent Interactions of Large Molecular Systems Parametrized for All spd-Block Elements ({Z} = 1–86)},
journal = {Journal of Chemical Theory and Computation},
volume = {13},
number = {5},
pages = {1989-2009},
year = {2017},
%doi = {10.1021/acs.jctc.7b00118},
%note ={PMID: 28418654},
}

@article{doi:10.1063/1.470117,
%author = {Essmann,Ulrich  and Perera,Lalith  and Berkowitz,Max L.  and Darden,Tom  and Lee,Hsing  and Pedersen,Lee G. },
author = {Essmann,Ulrich  and others},
title = {A smooth particle mesh Ewald method},
journal = {The Journal of Chemical Physics},
volume = {103},
number = {19},
pages = {8577-8593},
year = {1995},
%doi = {10.1063/1.470117},

}

@article{ricci2003algorithms,
  title={Algorithms for Brownian dynamics},
  author={Ricci, Andrea and Ciccotti, Giovanni},
  journal={Molecular Physics},
  volume={101},
  number={12},
  pages={1927--1931},
  year={2003},
  publisher={Taylor \& Francis}
}

@article{pulay1969ab,
  title={Ab initio calculation of force constants and equilibrium geometries in polyatomic molecules: I. Theory},
  author={Pulay, Peter},
  journal={Molecular Physics},
  volume={17},
  number={2},
  pages={197--204},
  year={1969},
  publisher={Taylor \& Francis}
}

@dataset{schade_robert_2021_4692508,
  %author       = {Schade, Robert and Kenter, Tobias and Elgabarty, Hossam and Lass, Michael and Schütt, Ole and Lazzaro, Alfio and Pabst, Hans and Mohr, Stephan and Hutter, Jürg and Kühne, Thomas D. and Plessl, Christian},
  author       = {Schade, Robert and others},
  title        = {{Enabling Electronic Structure-Based Ab-Initio 
                   Molecular Dynamics Simulations with Hundreds of
                   Millions of Atoms}},
  month        = apr,
  year         = 2021,
  publisher    = {Zenodo},
  %doi          = {10.5281/zenodo.4692508},
  url          = {https://doi.org/10.5281/zenodo.4692508}
}

@article{bowler2010calculations,
  title={Calculations for millions of atoms with density functional theory: linear scaling shows its potential},
  author={Bowler, David R and Miyazaki, T},
  journal={Journal of Physics: Condensed Matter},
  volume={22},
  number={7},
  pages={074207},
  year={2010},
  publisher={IOP Publishing}
}

@article{arita2014large,
  title={Large-scale DFT simulations with a linear-scaling DFT code CONQUEST on K-computer},
  author={Arita, Michiaki and Arapan, Sergiu and Bowler, David R and Miyazaki, Tsuyoshi},
  journal={Journal of Advanced Simulation in Science and Engineering},
  volume={1},
  number={1},
  pages={87--97},
  year={2014},
  publisher={Japan Society for Simulation Technology}
}

@ARTICLE{Kenney1991-sa,
  title     = "Rational iterative methods for the matrix sign function",
  author    = "Kenney, Charles and Laub, Alan J",
  journal   = "SIAM J. Matrix Anal. Appl.",
  publisher = "Society for Industrial \& Applied Mathematics (SIAM)",
  volume    =  12,
  number    =  2,
  pages     = "273--291",
  month     =  apr,
  year      =  1991
}

\end{document}